\documentclass[dvipsnames]{article}

\usepackage{amsmath}
\usepackage{mathtools}
\usepackage[ruled, norelsize]{algorithm2e}
\usepackage{amsthm}
\usepackage{amssymb}
\usepackage{esint}
\usepackage{nicefrac}
\usepackage{arydshln}
\usepackage{bm}
\usepackage{dsfont}

\usepackage{booktabs}
\usepackage{microtype}
\usepackage{graphicx}       
\usepackage{fancyhdr}   
\usepackage{verbatim}
\usepackage[]{hyperref}
\usepackage{url}
\usepackage{varwidth}
\usepackage{graphicx}
\usepackage{epstopdf}
\usepackage{float}
\usepackage{tikz}
\usepackage{setspace}
\usepackage{subcaption}
\usepackage{cleveref}
\usepackage[normalem]{ulem}
\usepackage{soul}      
\usepackage{etoolbox}
\usepackage{xcolor}
\usepackage{times}
\usepackage{comment}
\usepackage{multirow}

\usepackage{natbib}

\Crefname{algocf}{Algorithm}{Algorithms}

\DeclareMathOperator*{\argmin}{arg\,min}

\def \SVD{\operatorname{SVD}}
\def \PCA{\operatorname{PCA}}

\def \REJJOB{\operatorname{Reject-Job}}

\def \RSPCA{\operatorname{SSVD}}
\def \FPCA{\operatorname{Federated-PCA}}

\def \FPCAEC{\operatorname{FPCA-Edge}}

\def \energy{\operatorname{\mathcal{E}}}
\def \rankupdate{\operatorname{Rank}}
\def \QR{\operatorname{QR}}

\def \Y {\mathbf{Y}}
\def \R {\mathbb{R}}

\def \y{\mathbf{y}}

\newcommand{\B}[1]{\mathbf{#1}}

\def \rank {\operatorname{rank}}
 
\def \l{\left}
\def \r{\right}

\def \N{\mathbb{N}}

\def \wh{\widehat}

\newcommand{\CPUR}{\texttt{CPU Ready}\xspace}
\newcommand{\naive}{\texttt{naive}\xspace}
\newcommand{\Expsmo}{\texttt{ExpSmo}\xspace}
\newcommand{\ARIMA}{\texttt{ARIMA}\xspace}
\newcommand{\SVM}{\texttt{SVM}\xspace}

\DeclareMathOperator{\ind}{\mathds{1}}

\newcommand{\Pronto}{\textsc{Pronto}\xspace}%

\usepackage[accepted]{mlsys_style}

\begin{document}

\twocolumn[
\mlsystitle{Pronto: Federated Task Scheduling}

\mlsyssetsymbol{equal}{*}

\begin{mlsysauthorlist}
\mlsysauthor{Andreas Grammenos}{cst,at}
\mlsysauthor{Evangelia Kalyvianaki}{cst}
\mlsysauthor{Peter Pietzuch}{imp}
\end{mlsysauthorlist}

\mlsysaffiliation{cst}{Department of Computer Science and Technology, University of Cambridge, Cambridge, UK}
\mlsysaffiliation{at}{Alan Turing Institude, London, UK}
\mlsysaffiliation{imp}{Department of Computing, Imperial College London, London, UK}

\mlsyscorrespondingauthor{Andreas Grammenos}{ag926@cl.cam.ac.uk}

\mlsyskeywords{Machine Learning, MLSys}

\vskip 0.3in

\begin{abstract}

We present a federated, asynchronous, memory-limited algorithm for online task scheduling across large-scale networks of hundreds of workers.
This is achieved through recent advancements in federated edge computing that unlocks the ability to incrementally compute local model updates within each node separately. 
This local model is then used along with incoming data to generate a rejection signal which reflects the overall node responsiveness and if it is able to accept an incoming task without resulting in degraded performance.
Through this innovation, we allow each node to execute scheduling decisions on whether to accept an incoming job independently based on the workload seen thus far.
Further, using the aggregate of the iterates a global view of the system can be constructed, as needed, and could be used to produce a holistic perspective of the system.
We complement our findings, by an empirical evaluation on a large-scale real-world dataset of traces from a virtualized production data center that shows, while using limited-memory, that our algorithm exhibits state-of-the-art performance.
Concretely, it is able to predict changes in the system responsiveness ahead of time based on the industry standard \CPUR metric and, in turn, can lead to better scheduling decisions and overall utilization of the available resources.
Finally, in the absence of communication latency, it exhibits attractive horizontal scalability.

\end{abstract}
]

\printAffiliationsAndNotice{}

\section{Introduction}

Data center resource allocation or scheduling is the fundamental task of allocating resources (e.g., CPU, memory, network bandwidth, and disk space) to workloads such that their performance objectives are satisfied and the overall data center utilization is kept high.
Even small deviations from the desired objectives can have substantial detrimental effects with millions of dollars in revenue potentially lost~\citep{barrosoBook}. 

There exist many different data center scheduling approaches, e.g., ~\citep{apollo,mercury,borg,firmament,decima,sparrow}. Such approaches rely on estimates of nodes' future resource availability to schedule workloads in ways to avoid saturation and to efficiently utilize resources across data center nodes. 
For predicting resource availability, schedulers either probe available nodes on an on-demand basis~\citep{sparrow,borg}, or collectively analyze time-series of performance data generated by the underlying physical and virtual infrastructure (e.g., servers and virtual machines (VMs)) across data center nodes~\citep{cortez2017resource, decima}. 
For example, Microsoft’s Recourse Central gathers VM utilization data on a centralized cluster and uses offline machine learning prediction to tackle servers’ oversubscriptions~\citep{cortez2017resource}.
Although performance data (usually referred to as \textit{telemetry} data) can provide a very detailed view of resource consumption over time, it is a challenge how to effectively analyze this to accurately predict in \emph{near real-time} resource availability across nodes in a scalable manner. 

Related work has shown that when schedulers have access to performance data from all data center nodes, they canß generate improved holistic models for efficient provisioning~\citep{borg,firmament,apollo,decima, cortez2017resource}. 
However, these works operate on a near offline fashion and consume network bandwidth to transfer data from servers to centralized locations for processing. 
As such, they lack the ability to react to performance problems in real-time. 
Furthermore, to tackle data center scalability the vast majority of schedulers are distributed or hierarchical and continuously probe subsets of servers about their resource consumption, such as CPU and memory utilization, to make scheduling decisions based on nodes' availability after probing~\citep{sparrow}.
Although they can identify resource availability relatively fast they operate on a partial view of the data center and so they lack global efficiency.
It remains a challenge how to collectively analyze performance data in near real-time for efficient scheduling across data center nodes.  

To tackle the problem of large-scale performance data analysis with minimal latency we exploit recent advancements in edge computing.
Under this new paradigm, data is generated by commodity devices with potential hardware limitations and important restrictions on data-sharing and communication, which makes centralized processing of data extremely challenging.
The dominant, scalable training model that has shown to be able to tackle such challenges is Federated Learning (FL)~\citep{konevcny2016federated,mcmahan2016communication}.
Since the publication of these seminal works, we observe an increased interest in federated algorithms for training neural networks~\citep{smith2017federated, he2018cola, geyer2017differentially}.
In this setting, there is a large number of independent {\em clients} running at the edges that contribute to the training of a centralized model by computing local updates with their own data and sending them to a designated client holding the centralized model for aggregation ~\citep{yang2019federated}.
Over the year, FL has attracted significant interest and has eventually expanded into its own field with most of the literature to focus on deep neural networks, see~\citep{li2020federated}. 
The first notable example of a truly decentralized and federated cluster was put forth by Google to collaboratively train the Gboard\footnote{\url{https://ai.googleblog.com/2017/04/federated-learning-collaborative.html}} Android keyboard with great success.
Specifically, their approach used the technique from~\citep{konevcny2016federated} which was the first publicly announced {\em federated} method for training of neural networks. 
 
Despite the wide adoption of FL in many areas, to the best of our knowledge, this paradigm has not been yet applied into solving the large-scale, data analysis and data center task allocation problems.
Notably, traditional data centers are not federated, as all nodes operate under the same administration.
Here, we advocate that the need for real-time large-scale analysis of performance data makes FL the ideal solution for the problem in-hand.

We also believe that based on the trends observed~\citep{li2020federated,bonawitz2019towards} that federated schemes as the one used for training Gboard could expand in-scope and soon pave the way for the formation of \textit{federated data centers}. In this case, traditional schedulers which assume full access to all data center nodes will not be applicable. The scope of this paper is on exploiting FL for scheduling on today's data centers which could be further relevant to potentially future federated data centers.

In this paper, we present \Pronto, a federated algorithm that uses real-time performance data from virtualized data center nodes for online task scheduling.
Pronto, to the best of our knowledge, is the first to approach the data center scheduling problem as a federated processing one.
In particular, our approach focuses on predicting the spikes in values of the \CPUR performance metric generated by the VMware vSphere leading virtualization platform as these are generated by each data center node.
The \CPUR VMware metric captures the percentage of time a VM is ready to run but is not scheduled in one of the available CPUs. 
As a rule of thumb the \CPUR values should be kept low and below a predefined threshold for the system to operate well and without performance problems~\citep{davis2012demystifying}. 
Higher \CPUR values than the predefined threshold typically indicate that the VM’s performance is degrading as the VM in question does not run despite being ready~\citep{vsphereguide}. 
The \CPUR is used extensively as an industry standard indicator of performance problems~\citep{serverready}. 
Despite being an instrumental performance indicator with years of application in the industry, we are not aware of any data center schedulers based on this metric.

In \Pronto, we aim to predict the performance degradation of a node by detecting \CPUR spikes. 
At its core, \Pronto predicts an incoming \CPUR spike based on our empirical observation from a real-world data center trace that a spike in the weighted summation of the top-$r$ tracked projections within a node is highly indicative of an incoming spike.
To this end, \Pronto tracks in real-time the top-$r$ projections within each node and by exploiting spikes detected over a sliding window it decides whether to accept an incoming job or not.
Concretely, at each timestep and for every node, \Pronto generates a Boolean flag based on the weighted summation of the number of projection spikes detected over a sliding window. 
The flag is raised (\textit{i.e.}, \texttt{true}) if the node at time $t$ \textit{cannot} accept a job, and lowered (\textit{i.e.}, \texttt{false}) otherwise.
This Boolean flag over-time can be thought as a binary signal, which which we deb as the ``\textit{Rejection signal}''. 
In other words, we treat a system to be in a degraded state, if at any given time $t$ its \CPUR value exceeds a predefined threshold.

\Pronto is designed to be \textit{federated, streaming}, and \textit{unsupervised}. 
It is federated as it executes scheduling plans in a decentralized fashion without knowledge of the global performance dataset. 
This is one of the key benefits of our approach, as nodes are able to immediately take decisions about incoming workloads without the need for global synchronization and so reducing communication overhead and scheduling latency.
To achieve this, we combine recent advances of federated $\PCA$ to accurately compute the $d$-dimensional, $r$-rank embedding space $\mathbf{U}\in\R^{d \times r}$ of data incrementally~\citep{grammenos2019federated}. 
This approach also provides an efficient mechanism to adaptively estimate the embedding rank~\citep{grammenos2019federated} with high accuracy which is likely to happen as workload trends evolve.
Additionally, it is \textit{streaming} and only requires memory linear to the number of features considered, namely the required memory is proportional to $\mathcal{O}(d)$. 
Furthermore, it ensures \textit{data ownership}, as each node keeps its own incremental and evolving estimate while catering for distributional shifts; which can be crucial if some parts of the data center process orthogonal or sensitive workloads. 
Finally, experimental evaluation shows that it is \textit{fast} as it can incrementally track thousands of features per second while having no ``offline" components.
This means that the computation of \Pronto is \textit{streaming} requiring only a single pass over the incoming data without having to store any historical data in order to update its estimates.
This method can provide tangible benefits to data centers as it enables, in an online fashion, the allocation of incoming jobs across thousands of nodes efficiently and effectively while improving overall allocation.%
To the best of our knowledge this is the first work to tackle large-scale online workload scheduling using a FL-based algorithm. %

\textbf{The contributions of this paper are as follows.} { %
    We introduce a novel federated task scheduler that is streaming and 
    memory limited.
    It allows each of the computing nodes to independently take decisions for upcoming task assignments without the need for global synchronization.
    Further, each node only accepts an incoming job if its responsiveness will not be affected; a node's deterioration is captured by spikes of the \CPUR metric. 
    To the best of our knowledge, although \CPUR is widely used, it has not been used as a task scheduling predictor before.
    We also provide a thorough discussion about the importance of \CPUR for performance predictions and we present exploratory results on the use of traditional offline methods to predict \CPUR values. 
    Finally, we evaluate our proposed scheme using traces gathered from the virtualized data center of an international bank organization (hereafter referred to as the Company) to validate our claims. 
}

\section{Motivation}\label{sec:motivation}

The \CPUR is an important VM performance metric and it is widely used by system administrators to identify CPU saturation at-run-time. It reports the \% of time a VM is ready to run but is not scheduled in a CPU. The higher the \CPUR values are, the longer a VM is waiting for a virtual CPU (vCPU) to run and so its hosted workloads are not executing and suffer from performance degradation. When the \CPUR values of a VM are higher than a threshold then this VM is saturated and it needs more resources to run efficiently. 

Although the \CPUR metric is a key performance indicator, most schedulers focus on using the CPU utilization metric to allocate workloads to nodes. Typically, schedulers assume known CPU workload demands which they use to schedule workloads on node(s) where the predicted CPU availability is enough to satisfy the demands. To accommodate for mis-predictions on workload demands and nodes' availability and time-varying workload CPU utilizations, workloads are typically allocated with a higher than demanding share of nodes' CPU resources. Despite CPU over-subscription, workloads can still saturate nodes. The \CPUR metric is able to capture at real-time when workloads are inadequately provisioned with CPU resources. 

The empirical rule-of-thumb is to keep the \CPUR values below a small threshold number. This is shown by many online sources, e.g.,~\cite{cpuReadyOptvizor,cpuReadyVladan,cpuReadyIBM,cpuReadyLearnVMWare} and this is inline with private discussions we had with the Company's system administrators regarding the available to us dataset. The focus of this paper is to \emph{accurately} predict the \CPUR \textit{spikes} of a VM defined as the \CPUR values that exceed a certain threshold. The \CPUR spikes are caused by CPU saturation and so they can be used to identify CPU performance bottlenecks. 

\begin{figure}[htp]
    \centering
    \includegraphics[scale=.27]{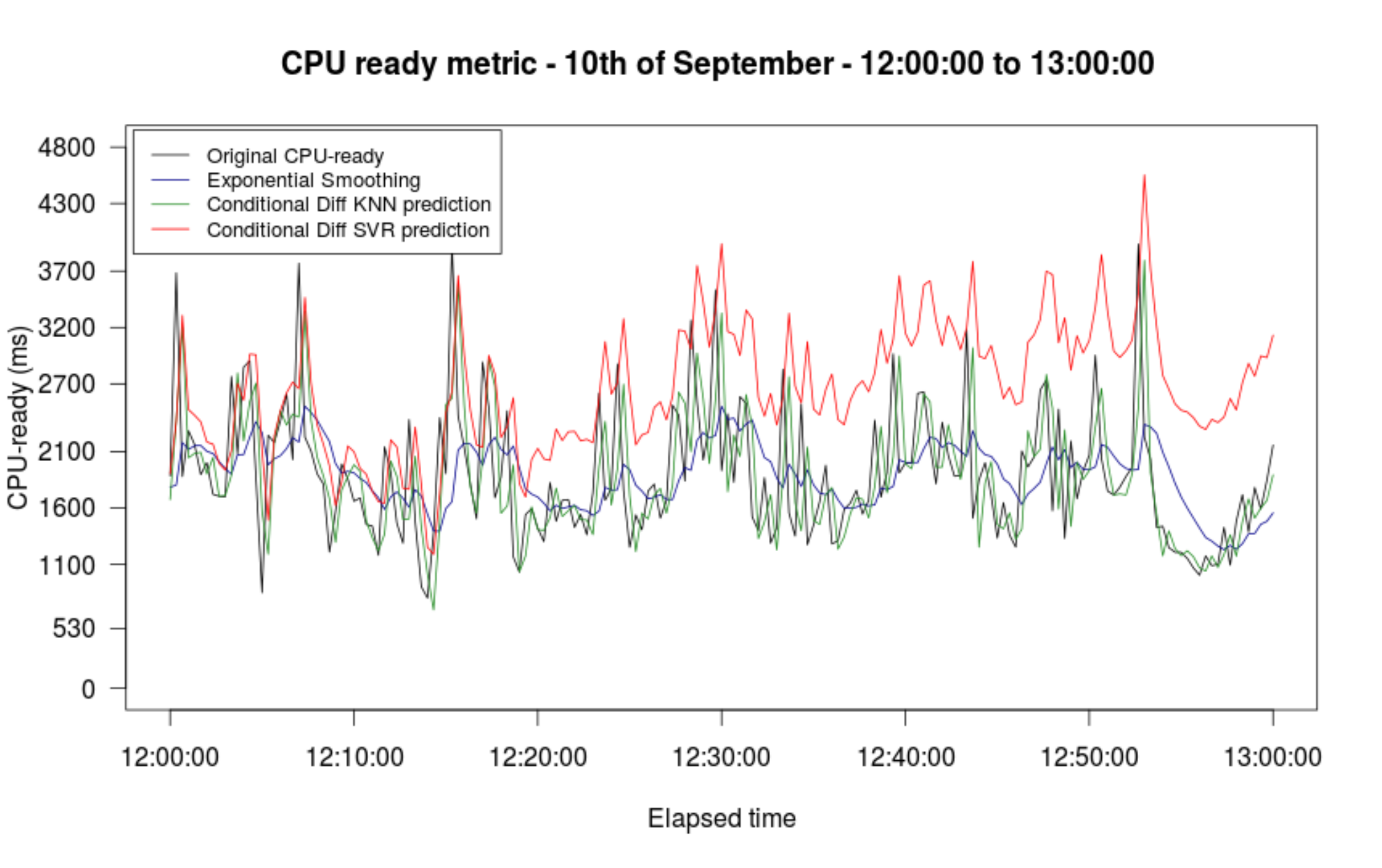}
    \caption{\CPUR real and predicted values for a single VM for one hour.}
    \label{fig:cpur-timeseries}
\end{figure}

Predicting the \CPUR values is challenging because: a) the \CPUR values are highly variable and do not follow regular patterns; b) spikes occur abruptly; and c) the values of the spikes vary significantly. For example, Figure~\ref{fig:cpur-timeseries} shows the \CPUR values for a single VM from our dataset for one hour. The \CPUR values are shown in ms which is the time the VM is ready to run but is not scheduled over a period of 20,000 ms. 

Accurately predicting the \CPUR values in~\Cref{fig:cpur-timeseries} is challenging. For instance, the same figure shows three different ways to predict the \CPUR values: the univariate exponential smoothing approach where the predicted value is a smoothed average of current values; and the multivariate approaches of  conditional Diff KNN and conditional Diff SVR which use additional performance metrics from the traces related to memory and network. All three methods work offline and use past measurements to predict future values. In particular, all three methods use a forecasting window of 20~secs based on training data from the last hour. Results show that none of these methods is able to accurately predict the \CPUR values. In the next section we perform a systematic exploration to investigating the accuracy of baseline and advanced forecasting methods for \CPUR prediction.

\section{\CPUR Prediction}\label{sec:prediction}

Our goal is to accurately predict the \CPUR spikes given past \CPUR values in order to identify future CPU contentions. We can predict \CPUR spikes using two approaches. The first approach forecasts \CPUR values and then uses the predicted values to find spikes above a certain threshold (Section~\ref{sec:pred:values}). The second approach focuses on predicting the \CPUR spikes directly using past spikes only (Section~\ref{sec:pred:spikes}). For both approaches we investigate the accuracy of various methods to predict the \CPUR values or spikes of a single VM and we pose the following fundamental questions: \textbf{(Q1)} \textit{how many other VMs do we need to look into?} \textbf{(Q2)} \textit{how much data from the past do we need to look into?} \textbf{(Q3)} \textit{how far in advance can we predict?} and \textbf{(Q4)} \textit{what is the best forecasting method to use?}

To evaluate our approach we use a real-world dataset of performance data gathered from the data centers of a global bank organisation from 2012. The dataset includes time-series VMware performance data from the Company's virtualized data centers. The Company collected such data across their virtualized infrastructure and stored it on NFS servers but derived little further value from it. The available to us dataset contains performance data from 100 virtualized clusters in a data center over four weeks. Each cluster has about 14 VMware ESX hosts, supporting 250–350 VMs. The performance data consists of metrics output by the VMware ESX hypervisor every 20 seconds related to the CPU, memory, disk and network resource consumption of physical hosts and VMs. There are 134 different resource metrics for a typical ESX host in our dataset, and 52 metrics for a VM. The total size of the dataset is 1TB.

\subsection{Predicting \CPUR values}\label{sec:pred:values}

First we forecast the per VM, daily, and median \CPUR values using past data from a) the same VM, b) all VMs belonging to same cluster, and c) all VMs with similar characteristics based on a pre-clustering phase. Predictions are calculated using windows of the past 14 and 21 days. Results are shown in Tables~\ref{tab:rmse:VM:VM} and ~\ref{tab:rmse:VM:kmeans} and aim to answer questions \textbf{Q1,} \textbf{Q2}, and \textbf{Q4}. Second, we report results of different forecasting windows of 1 day, 12, 6, 3 and 1 hour, and 30 and 15 mins using past windows of the same duration. Results are shown in Table~\ref{tab:rmse:VM:forecast} and aim to answer questions \textbf{Q3} and \textbf{Q4}.

\noindent{\bf Forecasting Methods:} We use the following forecasting methods: 1) \naive, where the prediction is the previous value seen per VM; 2) \texttt{exponential smoothing} (\Expsmo), per VM where the prediction is the smoothed average of the values in the current window and the newest values contribute more to the final output than the older ones. We obtain the best results setting $\alpha$ to 0.2; 3) \ARIMA (autoregressive integrated moving average), a popular time series forecasting approach. We tune the autoregressive, differencing and moving average (p,d,q) parameters locally for each forecast according to the smallest AIC criteria. This method is applied creating an average VM with all the VMs in the cluster used to build the \ARIMA model; 4) \SVM, an autoregressive transformation of the time series. We train an \SVM model for regression using the transformed data from all the VMs in the cluster. These four methods cover a wide spectrum of forecasting approaches from simple to more advanced approaches for robust results. 

Additionally, to improve the stability of the solvers and to systematize experimentation for all methods we normalize the input values to the [0,1] range (using simple scaling) according to the window size tested and then the values are ``de-normalized" before the error calculation. Evaluation is performed for three different VMs from three different clusters for the whole month of the trace. In this section we report the average Root Mean Square Error (RMSE) values for all three VMs.

\begin{table}[htb!]
    \centering
    \begin{tabular}{c|c|c|c|c}
    & \multicolumn{2}{c|}{same VM} & \multicolumn{2}{c}{same cluster VMs}  \\ \hline 
    method & 14 days & 21 days & 14 days & 21 days  \\ \hline \hline
    \naive   & 127.61 & 128.79   & 145.61    & 145.60 \\ \hline
    \Expsmo  & 127.04 & 129.71   & 145.59    & 148.77 \\ \hline
    \ARIMA   & 128.14 & 125.69   & 131.74    & 114.87 \\ \hline
    \SVM     & \bf{121.92} & \bf{118.01}   &\bf{103.66}    & \bf{100.23} %
    \end{tabular}
    \caption{Average RMSE when predicting \CPUR values per VM using data from the same VM and from same cluster VMs.}
     \label{tab:rmse:VM:VM}
\end{table}

\begin{table}[htb!]
    \centering
    \begin{tabular}{c|c|c}
    method & 14 days & 21 days \\ \hline \hline
    \texttt{Ordered} & \bf{102.62} & \bf{98.88} \\ \hline
    \texttt{KM Euclidean} & 106.33 & 102.42 \\ \hline
    \texttt{KM Corr} & 109.91 & 106.71 \\ \hline
    \texttt{KM Sts} & 112.13 & 105.59 \\ \hline 
    \texttt{KM Cort} & 113.03 & 107.10 \\ \hline 
    \texttt{KM Acf} & 104.31 & 102.02 \\ %
    \end{tabular}
    \caption{Average RMSE when predicting \CPUR values per VM using data from all ``similar" VMs based on KMeans (\texttt{KM}) clustering using different distance metrics shown in each row. Forecasting is performed using the SVM approach.}
     \label{tab:rmse:VM:kmeans}
\end{table}

Results in Table~\ref{tab:rmse:VM:VM} show that all methods and window sizes have significant errors. Different past window sizes do not significantly change the error in predictions.
Results from Table~\ref{tab:rmse:VM:VM} show that \ARIMA and \SVM forecasting tend to achieve a lower error. When comparing results between predictions using the ``same VM" and the ``same cluster VM" \ARIMA performs better when using data for larger windows in the case of ``same cluster VM". \SVM performs better regardless of the windows. Results from Table~\ref{tab:rmse:VM:kmeans} show that clustering-based forecasting can give better overall results when compared to Table~\ref{tab:rmse:VM:VM} but these results also show comparable performance to the \SVM approach from Table~\ref{tab:rmse:VM:VM}.

\begin{table*}[t!]
    \centering
    \begin{tabular}{c|c|c|c|c|c|c|c}
            & 1 day     & 12 hours  & 6 hours   & 3 hours   &  1 hour   & 30 min    &  15 min  \\ \hline \hline
    \naive   & 122.39    &   219.55  &   335.22  &   236.8   &   1052.41 & 837.71 &  876.16\\ \hline
    \Expsmo  & 123.42    &   183.78  &   273.81  &   269.08  &   899.66 &  798.21   &   933.69 \\ \hline
    \ARIMA   & 109.64    &   161.22  &   217.46  &   255.63  &   1154.43   & 1205.33 &   1515.90  \\ \hline
    \SVM cluster & 96.15 &   130.01  &   \bf{173.15} & \bf{155.58}   &   912.85 & 899.01    &   1155.12\\ \hline
    \end{tabular}
    \caption{Average RMSE when predicting \CPUR values per VM for different forecasting windows.}
     \label{tab:rmse:VM:forecast}
\end{table*}

Results from Table~\ref{tab:rmse:VM:forecast} show that the \texttt{SVM} method performs better for long forecasting windows i.e., 1 day, 12, 6 and 3 hours. For shorter windows of 1 hour, 30, and 15 mins the simple methods of \Expsmo and \naive give better results. Overall, as the forecasting window size decreases the RMSE gets substantially higher indicating how challenging it is to predict short term \CPUR values. 

To summarize and answer our initial questions we get better performance when we use data from multiple VMs (\textbf{Q1}), from a longer window from the past (\textbf{Q2}), and when we predict the average values for long forecasting windows (\textbf{Q3}). Finally, \SVM performs better across the different scenarios apart from the cases of short forecasting periods where the simpler \naive and \Expsmo methods perform better (\textbf{Q4}). However, and most importantly, all metrics show that none of these techniques can forecast the \CPUR values with high accuracy.

\subsection{Predicting \CPUR Spikes}
\label{sec:pred:spikes}

Rather than predicting the \CPUR values and then focusing on the predicted values above a certain threshold, in this section we predict \CPUR spikes directly. Here we present a baseline approach to predicting \CPUR spikes by transforming the time-series of \CPUR values into a new time-series of spikes and non-spikes and by using this new time-series we predict the spikes using the same methods as before. We call this approach the \texttt{alarm} method. 

We define a spike when the \CPUR value is above (or equals to) a given threshold. We define thresholds in the following ways. First, we define a \CPUR value to be a spike when this \CPUR value is above a certain \textit{fixed} threshold value. According to a quick inspection of the data we use the values of 500, 800, and 1000. Second, we use \textit{percentile}-based thresholds per VM. We use the values of $90^{th}$, $95^{th}$, and $99^{th}$ percentile. Third, we define per-VM statistically based thresholds. Assuming a normal distribution of the \CPUR values the threshold is computed as the average plus three times the standard deviation of a VM ($\mu+3\sigma$). We refer to this method as the \textit{statistical normal} threshold. Forth, we use a simplified xbar chart which calculates the upper control limit using a D4 correction over the moving range of the VM. We call this method as the \textit{statistical xbar} threshold. Finally, we also include results where the threshold is the \textit{median} of all values per-VM. 

Results are evaluated using the \textit{accuracy} error metric defined as: 
$\frac{\frac{\text{no of predicted spikes}}{\text{actual spikes}} + \frac{\text{no of predicted non-spikes}}{\text{actual non-spikes}}}{2}$. Results are shown in Tables~\ref{tab:accuracy:fixed},~\ref{tab:accuracy:percentiles}, and~\ref{tab:accuracy:stat} where the highest values are highlighted in bold and the last row of each table shows the \% of values considered as spikes by each method in the forecast set. 
The actual forecasting is performed as in~\Cref{sec:pred:values}. Results show the average values per-VM from three different clusters containing in total 1050 VMs. Predictions are performed for the next day.

\begin{table}[htb!]
    \centering
    \begin{tabular}{c|c|c|c}
    \hline
            & 500 & 800 & 1000  \\ \hline
    Naive   & 0.884 &   0.9359  &   0.9725   \\ \hline
    ExpSmo  & 0.8529 &  0.9338  &   0.9734  \\ \hline
    ARIMA   & 0.894 &   0.938   & \bf{0.9763}  \\ \hline
    SVM Cluster & 0.9057        & \bf{0.9417}   &   0.9746  \\ \hline
    SVM Full    & \bf{0.9071}   & 0.9414        &   0.9744  \\ \hline  \hline
     \% of spikes   & 9.54      & 2.63          &   0.85    \\ \hline  
    \end{tabular}
    \caption{Accuracy for spikes detection using a fixed threshold.}
     \label{tab:accuracy:fixed}
\end{table}

\begin{table}[htb!]
    \centering
    \begin{tabular}{|c|c|c|c|}
    \hline
            & $90^{th}$ & $95^{th}$ & $99^{th}$  \\ \hline
    Naive   & \bf{0.7472}   &   0.7807      &   0.8306  \\ \hline
    ExpSmo  & 0.7444        &  \bf{0.7942}  &  \bf{0.8534}  \\ \hline
    ARIMA   & 0.7301        & 0.7862        &   \bf{0.8534}  \\ \hline
    SVM Cluster & 0.745     &  0.7936       &   \bf{0.8534}  \\ \hline
    SVM Full    & 0.7449    &  0.7933       &   \bf{0.8534}  \\ \hline \hline
     \% of spikes   & 13.28      & 10.18          &   7.3    \\ \hline
    \end{tabular}
        \caption{Accuracy for spikes detection using percentiles.}
     \label{tab:accuracy:percentiles}
\end{table}

\begin{table}[htb!]
    \centering
    \begin{tabular}{|c|c|c|c|}
    \hline
            & $\mu+3\sigma$ & xbar  & median \\ \hline
    Naive   & 0.9723    & 0.6687    & 0.4476  \\ \hline
    ExpSmo  & \bf{0.9754}    & 0.5953    & 0.3393   \\ \hline
    ARIMA   & \bf{0.9754}    & 0.6594    & 0.4343  \\ \hline
    SVM Cluster & \bf{0.9754}    &  0.6875   & 0.4877  \\ \hline
    SVM Full    & \bf{0.9754}    &  \bf{0.6926}   & \bf{0.4903}  \\ \hline \hline
     \% of spikes   & 4.6      & 49.1          &   24.91    \\ \hline
    \end{tabular}
       \caption{Accuracy for spikes detection for the statistically-based and median methods.}
     \label{tab:accuracy:stat}
\end{table}

Results show that the \% of values considered as spikes varies widely depending on the method. This suggests that it is not easy to statically define a spike across methods. Regardless of the method used to determine spikes the \textit{accuracy} of forecasting these also varies widely amongst the spike detection methods. However, the best results in forecasting a certain type of spike mostly comes from the \texttt{SVM} approaches. Amongst all methods used the highest accuracy results are given by the methods which detect a small number of well defined spikes (i.e., value of 1000 in Table~\ref{tab:accuracy:fixed}, $99^{th}$ in Table~\ref{tab:accuracy:percentiles}, and ($\mu+3\sigma$) in Table~\ref{tab:accuracy:stat}. In the next sections we present the \Pronto approach to detect and predict \CPUR spikes using FL.

\section{Notation \& Preliminaries}
\label{label:prelim_notation}

This section introduces the notational conventions used throughout the paper. 
For integers $m\le n$  we use the shorthand $[m,n]=\{m,\cdots,n\}$ and $[n]$ for the special case $m=1$. 
If $\tau = Kb$ for some $K$ we let $[\tau]^{(b)}$ be the set of sets of the form $\{(k-1)b + 1, \dots, kb\}$ for $k \in [K]$. Whenever we use $[\tau]^{b}$ we will assume implicitly, and without loss of generality, that $\tau = Kb$. 
We use lowercase letters $y$ for scalars, bold lowercase letters $\mathbf{y}$ for vectors, bold capitals $\Y$ for matrices, and calligraphic capitals $\mathcal{Y}$ for subspaces.
If $\Y \in \R^{d \times n}$ and $S \subset [m]$, then $\Y_S$ is the block composed of columns indexed by $S$.
In particular, when $S=[k]$ for some $k \in \N$ we write $\Y_{[k]}$.
We reserve $\B{0}_{m\times n}$ for the zero matrix in $\R^{m \times n}$ and $\B{I}_{n}$ for the identity matrix in $\R^{n \times n}$.
Additionally, we use $\| \cdot \|_{\rm F}$ to denote the Frobenius norm operator and $\| \cdot \|$ to denote the $\ell_{2}$ norm.
If $\Y \in \R^{d \times n}$ we let $\mathbf{Y}=\mathbf{U}\mathbf{\Sigma} \mathbf{V}^T$ be its full $\SVD$ formed from unitary $\mathbf{U}\in \R^{d \times d}$ and $\mathbf{V} \in \R^{n\times n}$ and diagonal  $\mathbf{\Sigma} \in \R^{d \times n}$.
The values $\mathbf{\Sigma}_{i,i} = \sigma_{i}(\Y) \geq 0$ are the singular values of $\Y$.
If $1 \leq r\leq \min(d,n)$, we let 
 $\rho_r^2(\Y) = \sum_{i\ge r+1} \sigma_i^2(\Y)$ be the {\em residual} of $\Y \in \R^{m \times n}$ and 
$[\mathbf{U}_r, \mathbf{\Sigma}_r, \mathbf{V}_r^T] = \SVD_r(\mathbf{Y})$ be the singular value decomposition of its {\em best rank-$r$ approximation}. That is,
\[
 \mathbf{U}_r \mathbf{\Sigma}_r \mathbf{V}_r^T =  \argmin_{\mathbf{Z}\in \R^{d \times n}} \|\mathbf{Z}-\Y\|_F \;\text{ s.t. } \rank{(\mathbf{Z})} \le r.
\]
 Using this notation, we define $[\mathbf{U}_r, \mathbf{\Sigma}_r]$ to be the rank-$r$ {\em principal subspace} of $\Y$.
When there is no risk of confusion, we will abuse notation and use $\SVD_r(\mathbf{Y})$ to denote the rank-$r$ left principal subspace with the $r$ leading singular values $[\mathbf{U}_r, \mathbf{\Sigma}_r]$
We also let $[\B{Q}, \B{R}] = \mbox{QR}(\Y)$ be the QR factorisation of $\Y$ and $\lambda_{1}({\Y})\ge \cdots \ge \lambda_{k}(\Y)$ be its eigenvalues when $d=n$.
Finally, we let $\vec{\mathbf{e}}_{k} \in \R^d$ be the $k$-th canonical vector in $\R^d$.

\textbf{Streaming Model:} We define a data stream as a vector sequence $\mathbf{y}_{t_0}, \mathbf{y}_{t_1}, \mathbf{y}_{t_2}, \dots $ such that
$t_{i+1} > t_{i}$ for all $i \in \N$.
In this work, we assume that $\mathbf{y}_{t_j} \in \R^d$ and $t_j \in \mathbb{N}$ for all $j$.
Therefore, at time $n$, the data stream $\y_1, \dots, \y_n$
can be arranged in a matrix $\Y \in \R^{d \times n}$.
Streaming data models assume that, at each timestep, algorithms observe sub-sequences $\y_{t_1}, \dots, \y_{t_b}$ of the data rather than the full dataset $\Y$.

\textbf{Federated learning:}  
Federated Learning~\citep{konevcny2016federated} is a machine-learning paradigm that 
considers how a large number of {\em clients} 
owning different data-points can contribute to the training of a {\em centralized model} by 
locally computing updates with their own data and merging them to the centralized model without sharing data between each other.
Our method resembles the 
{\em distributed agglomerative summary model} (DASM)~\citep{tanenbaum2007distributed}
in which updates are aggregated in a ``bottom-up'' approach following a tree-structure.
That is, by arranging the nodes in a tree-like hierarchy such that, for any sub-tree, the leaves 
compute and propagate intermediate results for merging or summarisation.
In our model, the summaries only have to travel upwards once in order to be combined.
This permits minimal synchronization between the nodes so, for the purposes of this work, we do not model any 
issues related to synchronization.
We also note, that in the context of this work, we consider the terms ``job'' and ``task'' equivalent and are used interchangeably.

\section{A Federated Approach to Real-time Resource Monitoring} 
\label{sec:federated_scheduling}

An important element in task scheduling is knowing the resource availability of all nodes across the data center for globally informed allocation decisions. 
Maintaining such a global resource view is challenging and often centralized schedulers operate on cached data and distributed approaches work with different subsets of nodes to tackle large-scale scalability. 
To this end, we explore a different avenue for tackling this problem by exploiting recent advances on federated edge computing to \textit{accurately} track both the individual node status (``specialized view'') while also having the ability to get a holistic view of the system by intelligently synthesizing data from different nodes.
This approach allows each node to independently accept tasks, without the need for frequent global synchronization.

Formally, we consider a data center to be comprised out of $M$ nodes each having finite computing resources and each node has to accept the maximum number of jobs without impacting their own overall responsiveness.
Further, nodes produce a wealth of performance telemetry data used for scheduling, crash recovery, health monitoring, etc.
However, the number of such metrics that each node outputs can be highly dimensional, thus making their understanding and exploitation for scheduling in a streaming fashion very difficult.

We exploit unsupervised learning techniques as they are able to discover hidden correlations within unstructured data, with minimal assumptions about the input underlying structure.
Such methods, could eventually lead to improved allocation decisions with minimal user effort.
Out of the many techniques available, Principal Component Analysis  ($\PCA$)~\citep{pearson1901liii, jolliffe2011principal} is arguably the most ubiquitous one for discovering linear structure or reducing dimensionality in data, and so it has become an essential component in inference, machine-learning, and data-science pipelines. 
In a nutshell, given a matrix  $\Y \in \R^{d \times n}$ of $n$ feature vectors of dimension $d$, $\PCA$ aims to build a low-dimensional subspace of $\R^{d}$ that captures the directions of maximum variance in the data contained in $\Y$.
Each of these captured directions are $d$-dimensional vectors called Principal Components (PC's) which contain linear combinations of the features exhibiting their contribution to the total magnitude of each PC.
Conveniently, the PCs resulting from PCA are ordered in descending significance which is provided by the associated singular values.
However, even if the iterates can capture most of the information within the data they still lie in $d$-dimensional space. 
This high dimensionality makes their interpretation difficult and limits their usefulness.
To this end, and to effectively reduce its dimensionality, we exploit the resulting subspace estimate along with the incoming data in order to reveal the hidden patterns within the data.
These patterns can then be leveraged to improve our scheduling decisions.

The general intuition behind our scheme stems from the observation that each PC can be \textit{projected} to the incoming data, which yields a scalar value for each PC that can be tracked over time.
This value is a key indicator of the PC magnitude evolution over time and can reveal hidden trends and capture fluctuations of the overall direction pattern without knowing precisely which feature contributed to it at any given time.
Such trends and fluctuations could then be easily captured using traditional algorithms like averaging and Kalman filters.
This is because for each PC its projection is a single scalar that is tracked over time.

\begin{figure}[ht]
    \centering
    \includegraphics[width=\linewidth]{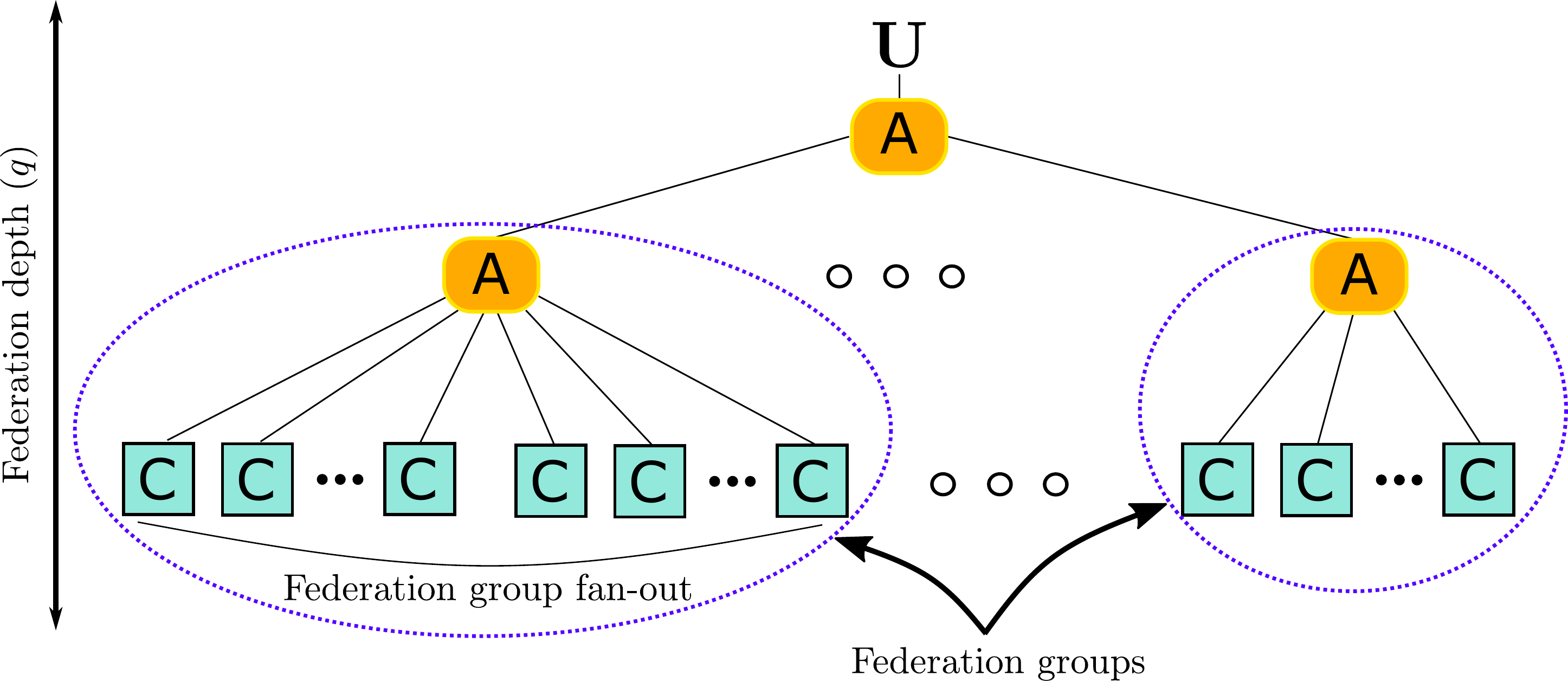}
    \caption{An indicative federation network structure in which compute nodes ($\mathbf{C}$) independently compute and make decisions based on the data they have seen so far. The updated subspaces are propagated upwards to aggregator nodes ($\mathbf{A}$) upon completion of every block and the difference between the previous subspace estimate is below $\epsilon$. Each subspace is propagated upwards until the root is reached. At this point we update the global estimate for the cluster workload seen thus far.}
    \label{fig:computation_graph}
\end{figure}

More formally, we assume that each of the compute nodes produces at each time-step a feature vector in $\R^{d}$ that contains all the metrics gathered. 
Each node can be thought as part of a decentralized dataset $\mathcal{D} = \{\y_{1}, \dots, \y_{n}\} \subset \R^d$ distributed across these $M$ clients.
The dataset $\mathcal{D}$ can be stored in a matrix
$\Y=\left[\Y^{1}|\Y^{2}|\cdots|\Y^{M}\right] \in \R^{d \times n}$ with $n \gg d$ and such that $\Y^{i} \in \R^{d \times n_i}$ is {\em wholly owned} by each compute client $i \in [M]$.
We assume that each $\Y^i$ is generated in a streaming fashion and that due to resource limitations it cannot be stored in full.
Furthermore, under the DASM we assume that the $M$ clients in the federation can be arranged in a computation graph with $q$ levels and computation nodes only as its ``leaves''. 
The computation graph contains computation nodes and aggregators which can be joined to create an independent federation group.
An example of such a computation graph is given in~\Cref{fig:computation_graph}.
We note that such structure can be generated easily and efficiently using various schemes~\citep{wohwe2019optimized}.
Moreover, we fully expect the computation graph to be shallow yet exhibit a very large fan-out which is typical for federated applications.

\subsection{Local Updates}

We now introduce our local update scheme which is responsible for producing the iterates we use within our scheduler. 
The local iterates are produced by exploiting $\FPCAEC$ which was put forth in~\citep{grammenos2019federated} and is also rank-adaptive. 
This is particularly convenient as it allows each of the processing nodes to adjust, independently of each other, their rank estimate based on the workload seen so far.
Its iterates are then used by our scheduler and play an important role for deciding at any given time if accepting a new job at the node has the potential to impact its performance. 
That is because these iterates contain a ``specialized'' view tailored for each node specifically, which reflects its overall responsiveness at any given time.
If the new job will severely impact the nodes' performance according to our metrics discussed below, the scheduler rejects the job, otherwise the job is accepted.

To elaborate on the local update algorithm internals, let us consider a sequence $\{\mathbf{y}_1, \dots, \mathbf{y}_n\} \subset \R^{d}$ of feature vectors and let their concatenation at time $\tau \leq n$ be 
\begin{equation}
\label{eq:conc of yts}
\Y_{[\tau]} = \l[ 
\begin{array}{cccc}
\mathbf{y}_1 & \mathbf{y}_2 & \cdots & \mathbf{y}_{\tau}
\end{array}
\r] \in\R^{d \times \tau}.
\end{equation} 
A block of size $b \in \N$ is formed by taking $b$ contiguous columns of $\Y_{[\tau]}$.
Hence, a matrix $\Y_{[\tau]}$ with $r \le b \le\tau$ induces $K = \lceil \tau/b \rceil$ blocks.
For convenience, we assume $K \in \N$, so that $\tau = Kb \in \N$. In this case, block $k \in [K]$ corresponds to the sub-matrix containing columns $S_k = \{(k-1)b+1, \dots, kb\}$.
It is assumed that all blocks $S_{k}$ are owned and observed exclusively by client $i \in [M]$, but that due to resource or time constraints can only store a small subset of them.
Hence, once client $i \in [M]$ has observed $\mathbf{Y}_{S_k} \in \R^{d \times b}$ it uses it to update its estimate of the $r$ principal components of $[\Y_{[(k-1)b]}, \mathbf{Y}_{S_k}]$.
If $\widehat{\Y}_{0,r}$ is the empty matrix, the $r$ principal components of $\Y_{[\tau]}$ can be estimated by computing the following iteration,
\begin{align}
[\B{U}_{k,r}, \B{\Sigma}_{k,r},\B{V}_{k,r}^T]&=\text{SVD}_{r}\left(\left[\begin{array}{cc}
\widehat{\Y}_{[(k-1)b],r} \,\,\, \Y_{S_k}\end{array}\right]\right)
\label{eq:svdr_1}\\
\widehat{\Y}_{[kb],r}&=\B{U}_{k,r} \B{\Sigma}_{k,r}\B{V}_{k,r}^T \in \R^{d \times kb}.
\label{eq:svdr_2}
\end{align}
Its output after $K$ iterations contains an estimate $\B{U}_{K,r}$ of the leading $r$ principal components of $\Y_{[\tau]}$ and the projection $\widehat{\Y}_{[\tau],r}=\B{U}_{K,r}\B{\Sigma}_{K,r}\B{V}_{K,r}^T$  of $\Y_{[\tau]}$ onto this estimate. 
Naturally, the closer $\widehat{\Y}_{[\tau],r}$ is to $\Y_{[\tau],r}$ the better our embedding.
Further, for every processed block each client keeps $r$ PC projections such that $\Y_{b,r}\B{U}_{k,r}\in\R^{r \times b}$ which allow us to track over the block duration if any anomalies (e.g. spikes) were found in any of the tracked PC's. 
The local subspace estimation we used is from~\citep{grammenos2019federated} which is also rank-adaptive; this allows each of the processing nodes to adjust, independently of each other, their rank estimate based on the workload seen so far.

\subsection{Global Updates}

In this section we describe how the global federation of the embedding happens in our scheme. 
Notably, the only data structure that is propagated upwards is the actual embedding and no nodes can perform predictions other than the leaf (computation) nodes themselves which happens in real-time and allows the scheduling to be both timely and independent.
To efficiently transfer knowledge of the workload embeddings seen so far, each node periodically requests a copy of the global embedding which can be merged against its local.
Further, this strategy can also be employed for new or transient nodes as they join into the computation pool.
Our algorithmic constructions are built upon the concept of {\em subspace merging} in which two subspaces $\mathcal{S}_1 =(\B{U}_{1}, \B{\Sigma}_1) \in \R^{r_{1} \times d} \times \R^{r_1 \times r_1}$ and $\mathcal{S}_2=(\B{U}_{2}, \B{\Sigma_2}) \in \R^{r_{2} \times d} \times \R^{r_2 \times r_2}$ are {\em merged} together to produce a subspace $\mathcal{S}=(\B{U}, \B{\Sigma}) \in \R^{r \times d} \times \R^{r \times r}$ describing the combined $r$ principal directions of $\mathcal{S}_1$ and $\mathcal{S}_2$.
One can merge two sub-spaces  by computing a truncated $\SVD$ on a concatenation of their bases. 
Namely,
\begin{equation}
    \label{eq:basic_subspace}
    [\B{U},\B{\Sigma},\B{V^{T}}] \leftarrow \SVD_{r}([\lambda \B{U}_{1}\B{\Sigma}_{1},\B{U}_{2}\B{\Sigma}_{2}]),
\end{equation}
where $\lambda \in (0,1]$ a {\em forgetting factor} that allocates less weight to the previous subspace $\B{U}_{1}$.
In~\citep{vrehuuvrek2011subspace,grammenos2019federated}, it is shown how \eqref{eq:basic_subspace} can be further optimized when $\mathbf{V}^{T}$ is not required and we have knowledge that $\mathbf{U}_{1}$ and $\mathbf{U}_{2}$ are already orthonormal, resulting in further resource savings. Due to spacing limitations the full implementation is deferred to the appendix.

\begin{figure*}
    \centering
    \includegraphics[width=\linewidth]{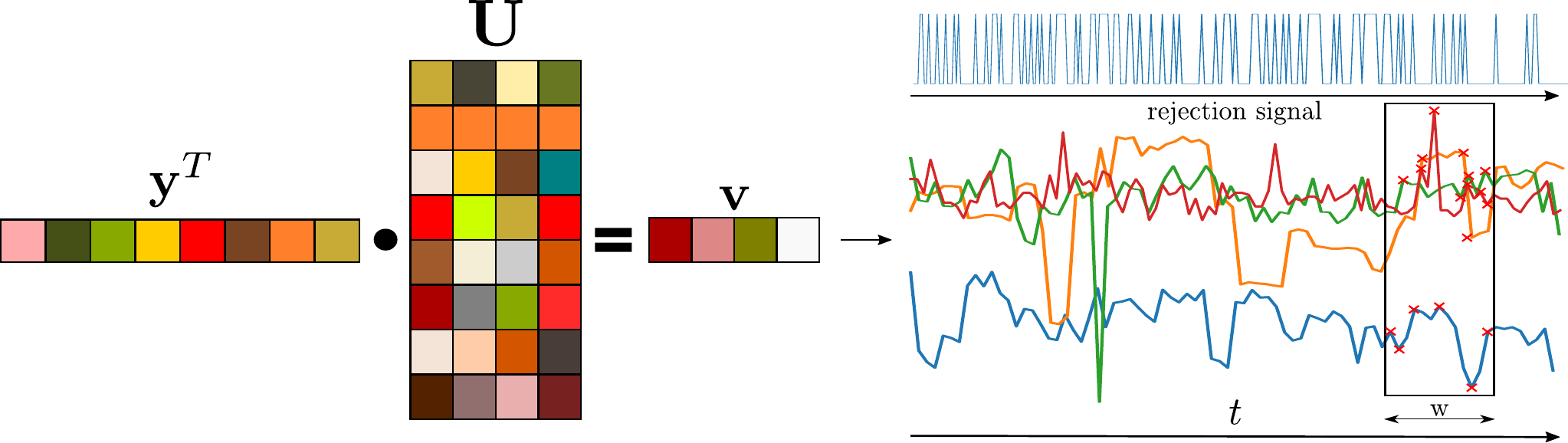}
    \caption{Projection of incoming $\B{y}\in\R^{d}$ onto embedding $\B{U}\in\R^{d \times r}$ producing $r$ projections in $\B{v}\in\R^{1\times r}$. Projections are tracked over time for detecting spikes which form the basis of our rejection signal. The sliding window for spike detection for each projection is of size $\B{w}$ also shown in the figure.}
    \label{fig:projection-process}
\end{figure*}

\section{\Pronto Scheduler} 

Herein, we propose a new task scheduler called \Pronto that is designed to accept an incoming job only if by doing so the performance of the existing running job(s) in the proposed node will not deteriorate significantly.
This deterioration is measured by spikes in \CPUR, as previously discussed.
Formally, at any given time $t$, \Pronto decides, if by accepting a new job at time $t$ will result in a \CPUR spike in the next few intervals after $t$.
In this section, we describe how by exploiting and extending the algorithmic constructions put forth in~\citep{grammenos2019federated}, we can derive a federated and reliable task cluster scheduler by exploiting the \CPUR metric.

The intuition behind \Pronto originates from our initial exploratory analysis and empirical observations on the available to us dataset that a spike in the top-$r$ tracked projections is indicative of incoming \CPUR spikes. 
Our proposed scheme exploits this observation by using it to compute a binary rejection signal based on the currently tracked embedding projections within each node. 
The rejection signal is raised if the weighted summation of the current nodes' tracked projections at time $t$ exceeds a predefined threshold.  
This event, indicates there enough projections spikes which implies rapid change in the magnitude of the tracked principal components.
Such rapid change is highly correlated with imminent \CPUR spikes, which as previously discussed, are highly indicative of degraded node performance.
\Cref{fig:projection-process} shows an overview of the system architecture for the scheduling process within each node.

\begin{figure}[ht]
    \centering
    \begin{subfigure}{.49\linewidth}
        \centering
        \includegraphics[width=\linewidth]{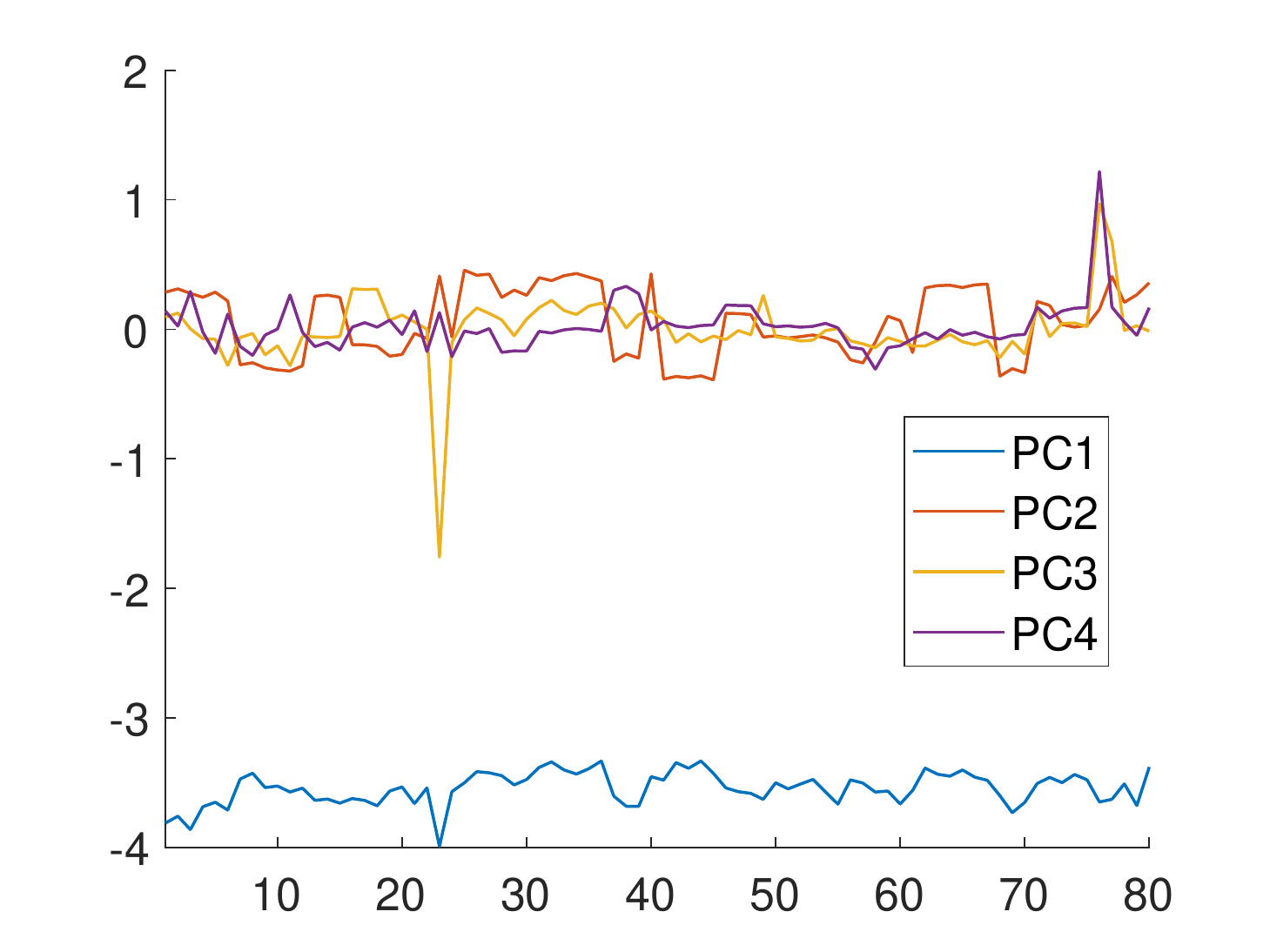}
        \caption{PC Projections}
        \vspace{11pt}
        \label{fig:projections-example}
    \end{subfigure}
    \begin{subfigure}{.49\linewidth}
        \centering
        \includegraphics[width=\linewidth]{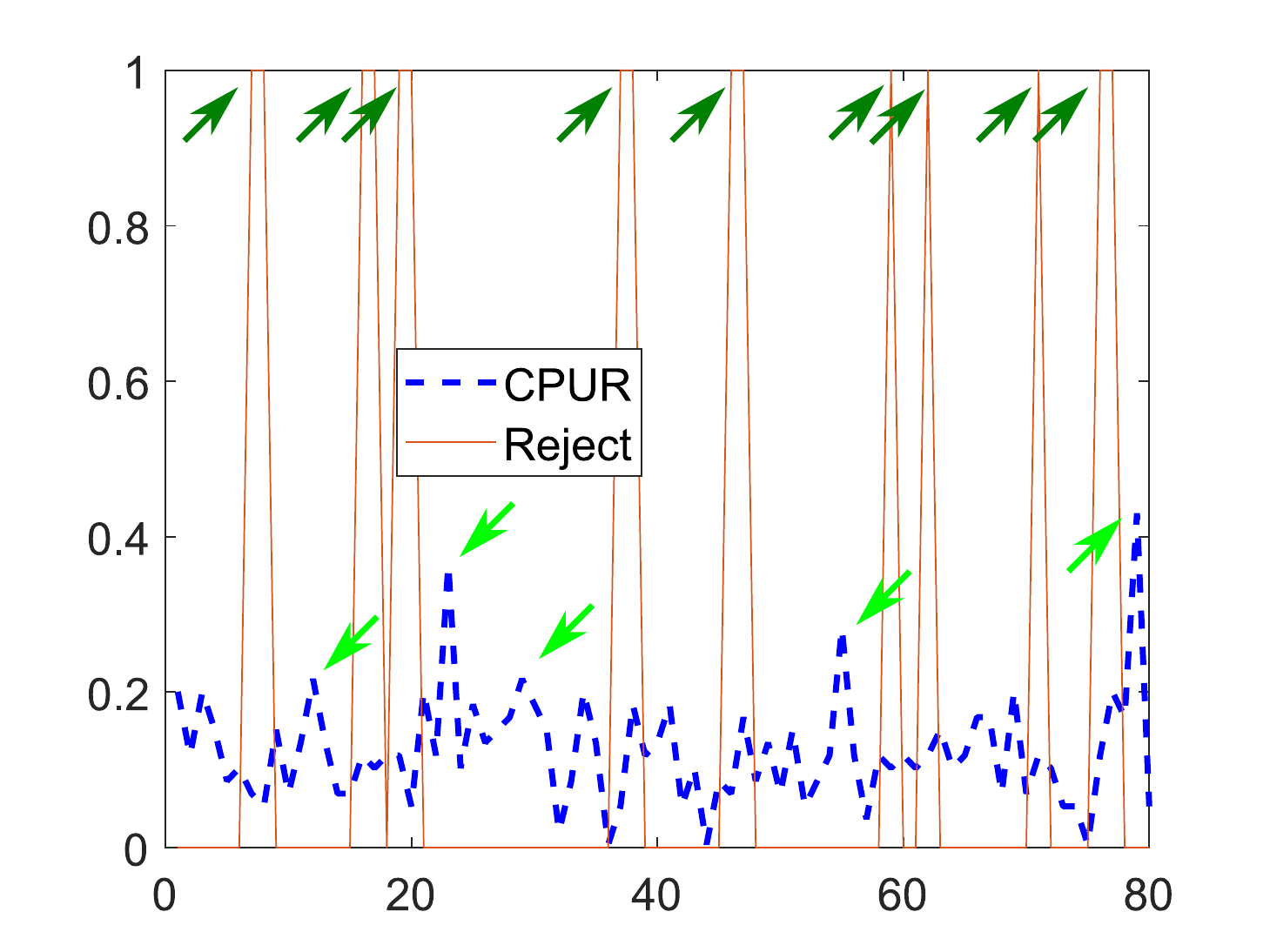}
        \caption{Rejection and CPU-Ready Signals}
        \label{fig:rejection-signal-example}
    \end{subfigure}
    \caption{
        Left~(\cref{fig:projections-example}): An example of the projections. Right~(\cref{fig:rejection-signal-example}): An example of the rejection signal based on the projections seen in the left drawn concurrently with the \CPUR signal serving as the ground-truth. 
        Further, \textit{green} and \textit{teal} arrows depict spikes in the rejection and \CPUR signals respectively. 
        Our goal is to show that spikes in the rejection signal \textit{precede} spikes in \CPUR over time.
    }
    \label{fig:cpur_example}
\end{figure}

To better illustrate the way the projections, the rejection signal, and the \CPUR signal work together consider~\Cref{fig:cpur_example}. 
At its left, we can see the tracked projections over time for a particular node, for which we see various rapid changes to them over time. 
On the right, we observe the rejection signal over the same period of time along with the detected spikes of \CPUR.
In this instance the spike threshold for \CPUR is set to be below $.2$ and the rejection signal is produced by observing the spikes in an online fashion of the projections shown in~\Cref{fig:projections-example}.
Provided with this, we can observe that each of the detected \CPUR spikes is \textit{preceded} by at least one raise of the rejection signal within few timesteps of its occurrence. 
Note, that consecutive \CPUR spikes might indicate that for the next few intervals the node will experience deteriorated performance and thus we cannot possibly accept a job - which is precisely what \Pronto exploits.

Elaborating, each node at a given time $t$ uses its own subspace iterate $\B{U}\in\R^{d \times r}$ as produced by $\FPCAEC$ and each incoming vector of features in $\R^{d}$ is projected into it to produce the projections of that node at time $t$.
As mentioned previously for each of the tracked projections, we monitor their abrupt changes (\textit{i.e.}, the signal spikes) in a streaming fashion over a sliding window of size $w$.
Our spike detection algorithm is based on a z-score based scheme put forth in~\citep{vanBrakel2014streamingpeak} which we implement within our rejection signal computation.
The results of the detection are stored as $r$ binary variables setting each to $1$ if a positive and to $-1$ if a negative spike is detected for that projection at time $t$ while being $0$ otherwise.
Moreover, at each time $t$ we compute the weighted sum of these $r$ binary variables multiplied by their associated singular value as, $R_{s}=\sum_{i=0}^{r}r_{i,t}\sigma_{r_{i},t}$\label{rej_weighted_sum}
where $r_{i,t}$ and $\sigma_{i,t}$ the $i$-th binary variable and $i$-th singular value at time $t$ respectively.
Then, the rejection signal value at time $t$ is set to $1$ if the weighted sum of $R_{s}$ is above a certain threshold and $0$ otherwise; throughout our experiments we set the threshold value to be equal to $1$.
We present the implementation of $\REJJOB$ in~\Cref{algorithm:reject-job}.

\begin{algorithm}[ht]
\scriptsize{
    \DontPrintSemicolon
    \SetNoFillComment
    \SetKwProg{Fn}{Function}{ is}{end}
    \KwData{$\B{U}\in\R^{d \times r}$: embedding estimate at time $t$, $\B{\Sigma}\in\R^{r}$: embedding singular values at time $t$, $\y\in\R^{d}$: observed data-point at time $t$, $\B{w_{p}}\in\R^{r \times \textrm{lag}}$: the dampened signal, $\B{w}_{\mathrm{avg}}\in\R^{r \times \textrm{lag}}$: the average filter, $\B{w}_{\mathrm{std}}\in\R^{r \times \textrm{lag}}$: the std filter
    }
    \KwResult{True if a job should be rejected at $t$, false otherwise.}
    \Fn{$\REJJOB(\B{U}, \B{\Sigma}, \y, \B{w_{p}}, \B{w}_{\mathrm{avg}},\B{w}_{\mathrm{std}})$}{
        \tcc{Init.}
        lag $\leftarrow 10$,  $\alpha \leftarrow 3.5$, $\beta \leftarrow 0.5$, $\B{b} \leftarrow 0_{1 \times r}$, $\textrm{tr} \leftarrow 1$\;
        \tcc{Compute projections}
        $\B{p} \leftarrow \y^{T}\B{U}\in\R^{1 \times r}$\;
        \tcc{Has lag buffer filled?}
        \If{$\mathrm{len}(\B{w}_{\mathrm{avg}} < \mathrm{lag}$)}
        {\Return false}
        \tcc{Find spikes for each projection}
        \For{$i=0;i<\mathrm{len}(\B{p});i++$}{
            \eIf{$\mathrm{abs}(\B{p}[i] - \B{w}_{\mathrm{avg}}[i][\mathrm{lag}-1]) > \alpha\B{w}_{\mathrm{std}}[i][\mathrm{lag}-1]$}{
                \tcc{Peak detected, check sign}
                \eIf{$\B{p}[i] > \B{w}_{\mathrm{avg}}[i][\mathrm{lag}-1])$}{
                    $\B{b}[i] \leftarrow 1$
                }
                {
                $\B{b}[i] \leftarrow -1$\;
                \tcc{Reduce $\B{w_{p}}$ influence}
                $\B{w_{p}}[i][\mathrm{lag}] \leftarrow \beta \B{w_{p}}[i][\mathrm{lag}] + (1-\beta)\B{w_{p}}[i][\mathrm{lag}-1]$
                }
            }
            {
            \tcc{No spikes}
            $\B{b}[i] \leftarrow 0$\;
            $\B{w_{p}}[i][\mathrm{lag}] \leftarrow \B{p}[i]$
            }
            \tcc{Adjust the buffers}
            $\B{w}_{\mathrm{avg}}[i,:] \leftarrow \mathrm{mean}([\B{w}_{\mathrm{avg}}[i, 2:], \B{p}[i]])$\;
            $\B{w}_{\mathrm{std}}[i,:] \leftarrow \mathrm{std}([\B{w}_{\mathrm{std}}[i, 2:], \B{p}[i]])$
        }
        \tcc{Do we reject a job?}
        \eIf{$\sum_{i}(\B{b}[i]\B{\Sigma}[i])$ greater or equal than $\mathrm{tr}$ threshold}
        {\Return true}
        {\Return false}
    }
}
    \caption{Reject-Job} 
    \label{algorithm:reject-job}
\end{algorithm}

\noindent This principled approach in predicting \CPUR spikes is enabled due to the properties inherited by the incremental embedding computation as described in~\citep{grammenos2019federated} and specifically~\Cref{alg:fpca_edge} in the appendix. 
Exploiting the algorithm introduced previously, not only provides us with concrete guarantees with respect to the embedding quality but is also systematic and deterministic with the only variable left to tune being the threshold for raising the rejection signal.
Note, that in this instance we do not require the differential privacy aspects provided by $\FPCAEC$ and thus the perturbation masks are disabled throughout.

\begin{figure}[ht]
    \centering
    \includegraphics[width=\linewidth]{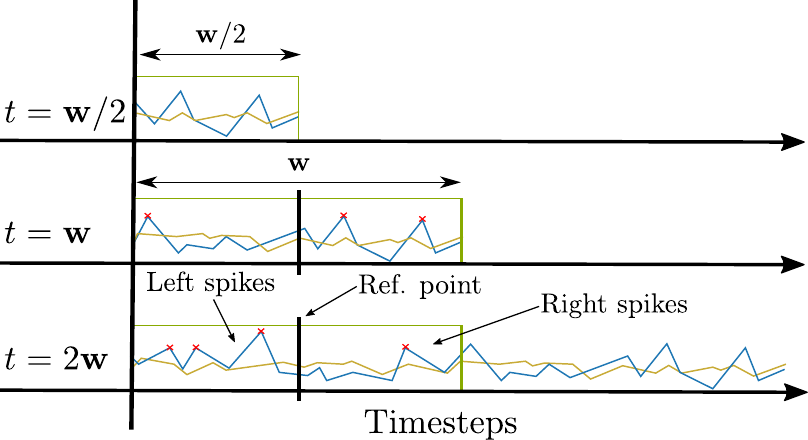}
    \caption{
        Illustration using two projection signals over three indicative timesteps, namely $t=\{\mathbf{w}/2, \mathbf{w}, 2\mathbf{w}\}$.
        It shows how the sliding window operates and indicates that the minimum observations required to start predicting is equal to the size of our window $\mathbf{w}$.
        In the final row we show where left- and right-sided spikes reside with respect to the reference point, which also acts as the current time for the scheduler.
        These spikes for each projection - after $\mathbf{w}$ observations - are then used to compute the rejection signal as described previously.
    }
    \label{fig:left_right_spikes_illustration}
\end{figure}

In order to see how the spikes are detected in practice, we can observe what happens to a node upon initialization.
This is depicted in~\Cref{fig:left_right_spikes_illustration} which shows three indicative snapshots at different timesteps that illustrate how spike detection works in practice.
As we previously said we cannot start making predictions until at least $\mathbf{w}$ observations have been processed.
At that point, we set out a reference point through which we set our prediction horizon.
This is shown in the first row of the figure.
In the second row of the same figure we can see that a full window has been observed and thus we can start to reliably detect potential spikes.
This relative point is essentially what \Pronto considers its current time. 
This means that the spikes happen to the left of the reference point are considered to be incoming predictions while the ones on the right side are considered to have already happened in the past.
This segmentation is shown in the third row of the aforementioned figure.

Having fully described how the rejection signal is computed, we now describe our federated scheduling scheme.
To do so, we will exploit the $\FPCA$ technique introduced in~\citep{grammenos2019federated}.
Practically, we aim to establish a principled way that the iterates can be propagated upwards in order to be able to generate a holistic view of the system exactly like the way $\FPCA$ is designed to do.
We now present our \Pronto scheduler implementation depicted in~\Cref{algorithm:pronto-rejection-scheduler}.

\begin{algorithm}[htb!]
\scriptsize{
    \DontPrintSemicolon
    \SetNoFillComment
    \SetKwProg{Fn}{Function}{ is}{end}
    \SetKwProg{Each}{Each client}{ :}{end}
    \SetKwProg{At}{At time}{ , each client $i \in [M]$ with job set $[J]$}{end}
    \KwData{%
    $\Y=\left[\Y^{1}|\cdots|\Y^{M}\right] \in \R^{d \times n}$: Data vectors\;
    $(\alpha, \beta)$: Upper and lower bounds on $\energy(\Y)$\;
    $b$: Batch size at the client level\;
    }
    \Fn{$\text{Pronto}(\Y, b\mid \alpha, \beta)$}{
        \tcc{Initialise clients}
        \Each{ $i\in [M]$}{
            Initialises PC estimate to $(\B{U}^i, \B{\Sigma}^i) \leftarrow (0,0)$\;
            Initialises batch to $\B{B}^i \leftarrow \left[\;\right]$\;
        }
        Arrange clients' merge paths in a computation graph such as in~\autoref{fig:computation_graph}\;
        \tcc{Computation of local updates}
        \At{$t \in [T]$}{
            Observes data-point $\y^i_{t} \in \R^d$\; 
            Adds $\y^i_t$ to batch $\B{B}^i \leftarrow [\B{B}^i, \y^i_{t}]$\;
            Accepts job $j'\in[J]$ based on $\REJJOB$\;
            \If{$\B{B}^i$ has $b$ columns}{
                $(\B{U}^i, \B{\Sigma}^i) \leftarrow \FPCAEC$\; %
                $\B{B}^i \leftarrow \left[\;\right]$\;
            }
        }
        \tcc{Merge only if needed}
        \If{$\mathrm{absdiff}(U^{t,i}, U^{t-1,i})>\epsilon$}{
        Use~\Cref{algorithm:fastest_subspace} to merge subspaces recursively within the computation graph (Fig.~\ref{fig:computation_graph})\;
        }
    }
}
    \caption{Pronto scheduler}
    \label{algorithm:pronto-rejection-scheduler}
\end{algorithm}

\noindent Note, that in our current setting and in order to save bandwidth we can elect to propagate only if the estimate has changed above a certain threshold. 
To achieve this, we employ a heuristic to check if the absolute weights of the subspace iterate surpassed the set threshold.

\section{Evaluation}
\label{sec:eval}

This section is focused on the empirical evaluation of \Pronto against real-world traces. 
In a nutshell, the primary goal of this evaluation is to quantify the efficiency of our scheme to predict the \CPUR spikes based only on the Company's unstructured trace observed from each of the compute nodes. 

Formally, we use \Pronto to predict at any given time $t$ that a raise in the rejection signal occurs shortly before or coincides with an observed \CPUR spike contained in a sliding window of size $w$.
This enables us to accurately predict incoming \CPUR spikes which we can use to prevent further accepting additional jobs on a node with performance shortages.
We use the Company's dataset trace to evaluate our approach.
The rejection signal is then compared with the baseline (i.e., the actual \CPUR values) and we check if a spike is indeed detected in \CPUR and if that's reflected in the rejection signal.
If the rejection signal was \textit{raised} a few timesteps before or coincides with the actual \CPUR spike we classify it as a successful prediction. 
As noted in the previous section, the window $w$ of timesteps can be easily adjusted. 
However, throughout our evaluation we observe that values close to ten timesteps give us good performance so we use this value for all experiments.

Practically speaking, to generate the rejection signal we only need the actual embedding ($\B{U}$) and its singular values ($\B{\Sigma}$). 
The actual subspace embedding $\B{U}$ can be generated using a plethora of methods which we also evaluate.
To this end, we evaluate against the one provided by~\citep{grammenos2019federated} and the methods of  SPIRIT~\citep{papadimitriou2005streaming} (SP), frequent directions~\citep{liberty2013simple} (FD), and finally power method~\citep{mitliagkas2013memory} (PM).
These algorithms represent the state-of-art in the area, are well-established, and each uses a different mathematical approach into tackling this problem.
Unfortunately, none of these algorithms are neither inherently distributed and, apart from SP, rank-adaptive.
We note however, that the singular values are needed for \Pronto to work in a truly federated setting which can only be reliably produced by~\citep{grammenos2019federated} and partially by~\citep{papadimitriou2005streaming}.
Concretely, SP is able to produce singular values but without any guarantees about their quality, while both FD and PM lack the ability to produce any. 
For the methods that are not able to generate their own singular values we use predefined values  generated using an exponential decay spectrum, namely $\sigma_{r}= 1/r$.
While not exact, this approximation enables us to test against all of the competing methods.

\subsection{Simulation Results}   

As we discussed before, our goal is to evaluate how efficiently incoming \CPUR spikes can be predicted and so we use the actual Company's trace as our baseline.
\Pronto is designed to operate using a sliding window of size $w$ for which we use the $\REJJOB$ algorithm to decide if at time $t$ the node can accept an incoming job or not.
In reality, this sliding window imposes a slight ``lag'' between the observed values and actual prediction time when the scheduler has to decide. 
We elect to at least see one window in order to make a prediction which equates in our case to $w$ timesteps throughout our experiments.
We believe this delay will, in practice, be insignificant as the rate of incoming trace observation is much higher than the number of incoming jobs.
Note also that typical window sizes in practical applications should range between $10-50$ timesteps\footnote{This applies to \Pronto itself, SPIRIT, and FD. PM needs to have a block size at least equal to the dimensionality of the data, which necessitates a larger window than the other methods.}.
Further, throughout our experiments we use a rank $r$ equal to $4$. Evaluation on higher values provided little to no benefit in terms of prediction quality with the added downside of increasing computation cost.

Before presenting the results we need to describe how we define a successful prediction.
Hence, in this context we classify a successful prediction if a \CPUR spike is preceded by \textit{at least} one rejection signal raise within the current window.
Moreover, we use the reference point within the window which equals half of the window size, namely $\mathbf{w}/2$. 
Thus, at any given time we can classify the detected spikes into \textit{left} and \textit{right} based on their location in the current window relative to our reference point. 

Formally, we measure at each timestep $t$ how many spikes are contained within the current sliding window as well as their ``side'' with respect to the reference point $t$.
We also measure the overall downtime, which is the amount of time the rejection signal is raised during our evaluation.
This reflects the overall job acceptance availability at each node.
Ideally, we would like to have the rejection signal raised before or coincide with a \CPUR spike, but also minimize the downtime so that the node can accept more jobs.

\begin{figure}[ht!]
    \centering
    \begin{subfigure}{.49\linewidth}
        \centering
        \includegraphics
        [width=\linewidth]
        {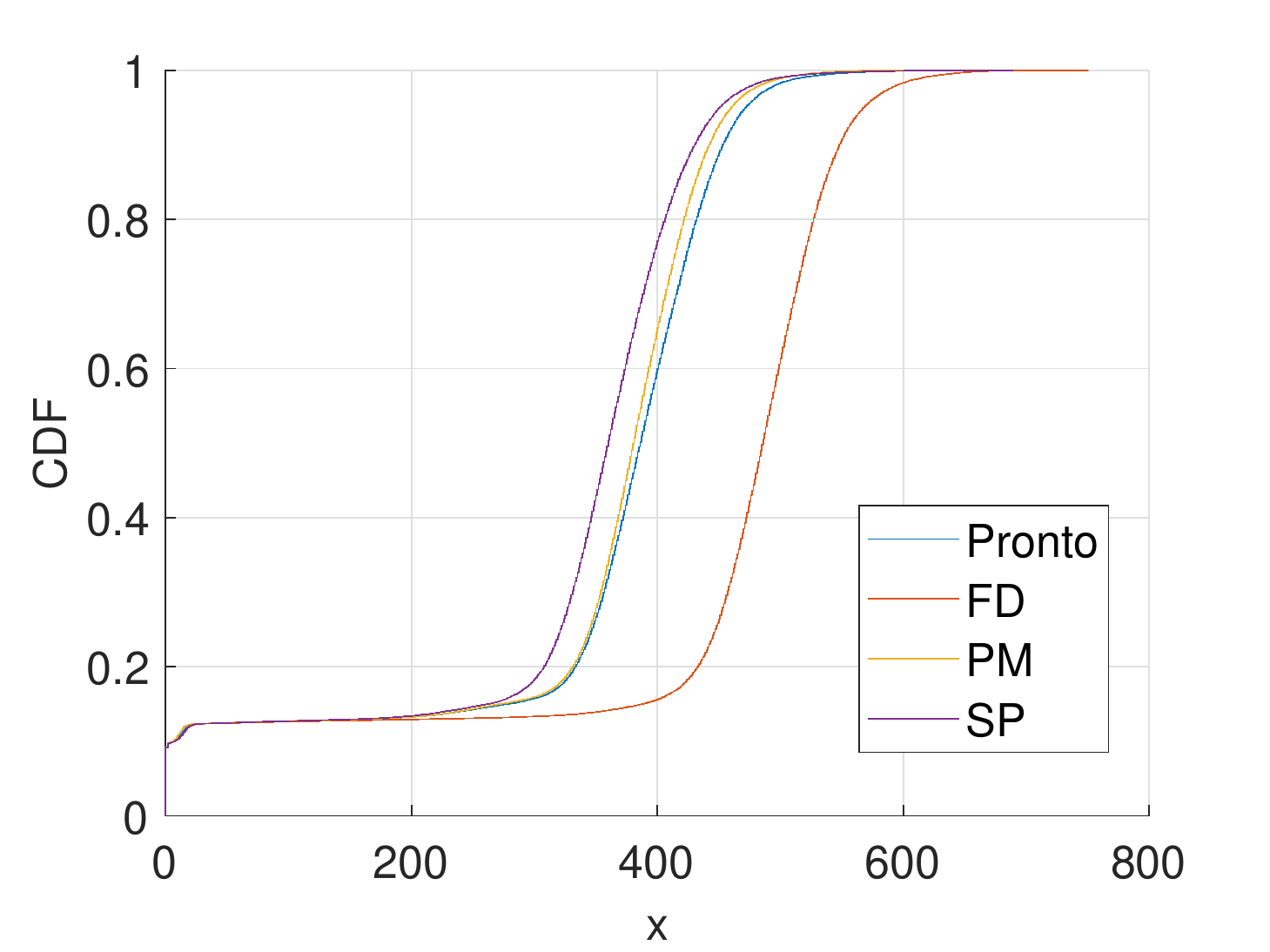}
        \caption{\textit{Left-sided} spikes CDF}
        \vspace{11pt}
        \label{fig:left-sided-peaks-cdf}
    \end{subfigure}
    \begin{subfigure}{.49\linewidth}
        \centering
        \includegraphics
        [width=\linewidth]
        {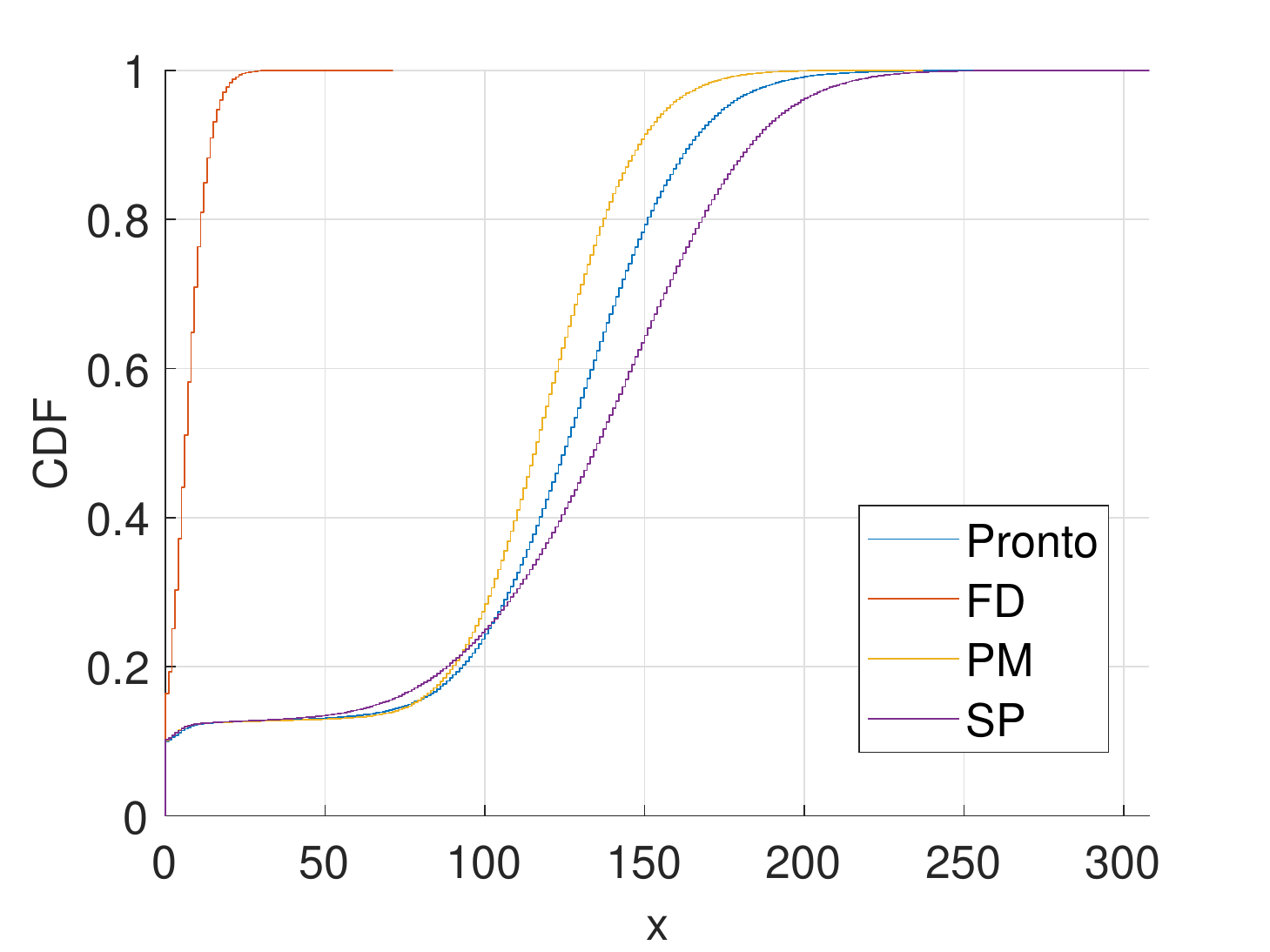}
        \caption{\textit{Right-sided} spikes CDF}
        \vspace{11pt}
        \label{fig:right-sided-peaks-cdf}
    \end{subfigure}
    \caption{
        Left: The empirical CDF of the number of \textit{left-sided} spikes. 
        Right: The empirical CDF of the \textit{right-sided} spikes. 
        Both are measured against our reference point at time $t$ with respect to \CPUR spike.
    }
\end{figure}

We start by quantifying the spike behaviour and we evaluate the types of spikes observed throughout the trace and the empirical Cumulative Distribution Functions (CDF) are presented in~\Cref{fig:left-sided-peaks-cdf} and~\Cref{fig:right-sided-peaks-cdf} for \textit{left-} and \textit{right-sided} spikes respectively.
We note, that \textit{left-sided} spikes are the most important ones, as these indicate that a \CPUR spike is imminent in the next few timesteps.
\textit{Right-sided} spikes can be an indication of consecutive \CPUR spikes or delayed detection; this is because \CPUR spikes might occur consecutively or very close to each other thus indicating a significantly deteriorated node. %

\begin{figure}
    \centering
    \begin{subfigure}{.49\linewidth}
        \centering
        \includegraphics[scale=.27]{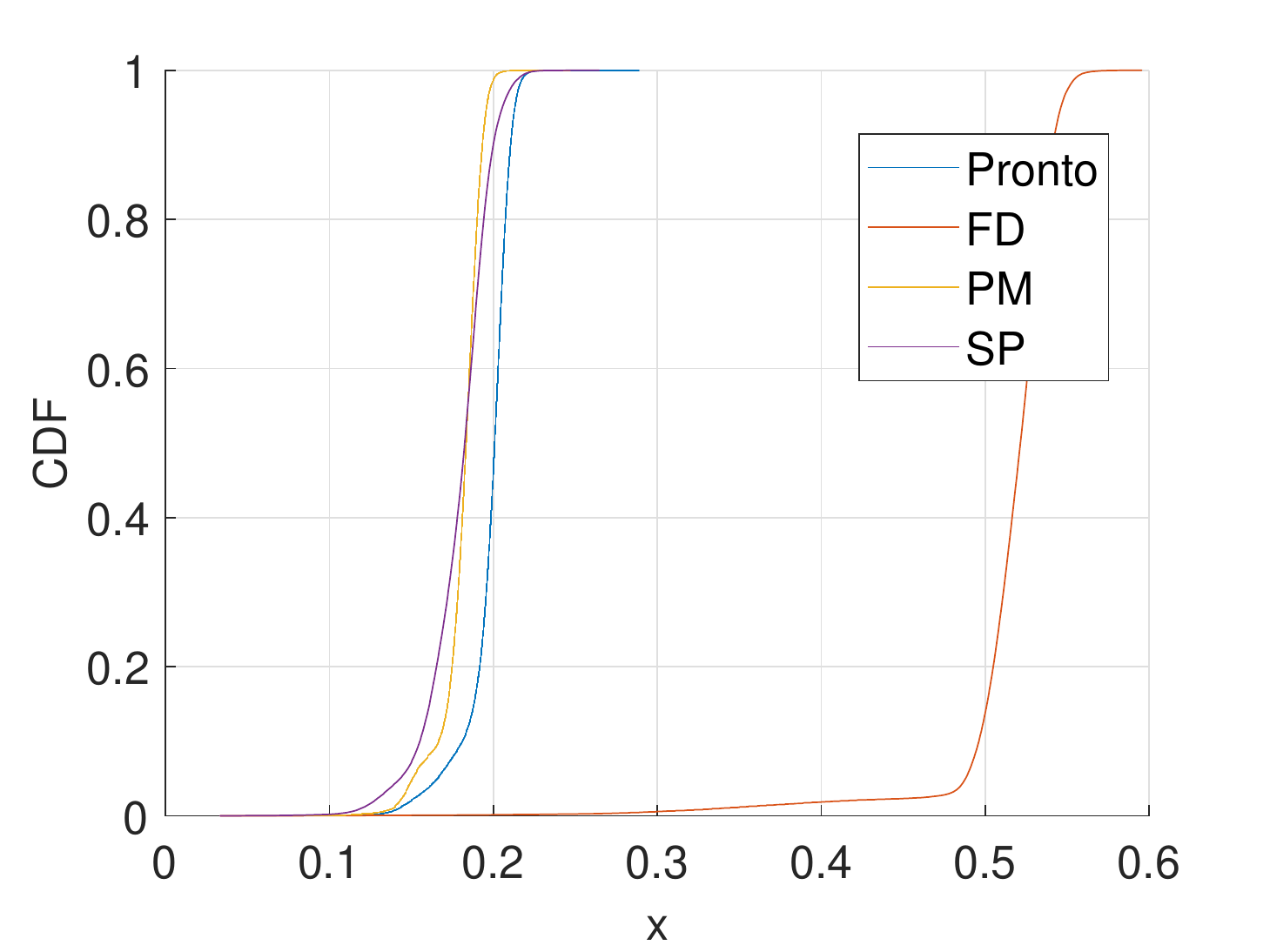}
        \caption{CDF of the percentage of time that the rejection signal was raised}
        \label{fig:downtime-cdf}
    \end{subfigure}
    \begin{subfigure}{.49\linewidth}
        \centering
        \includegraphics[scale=.27]{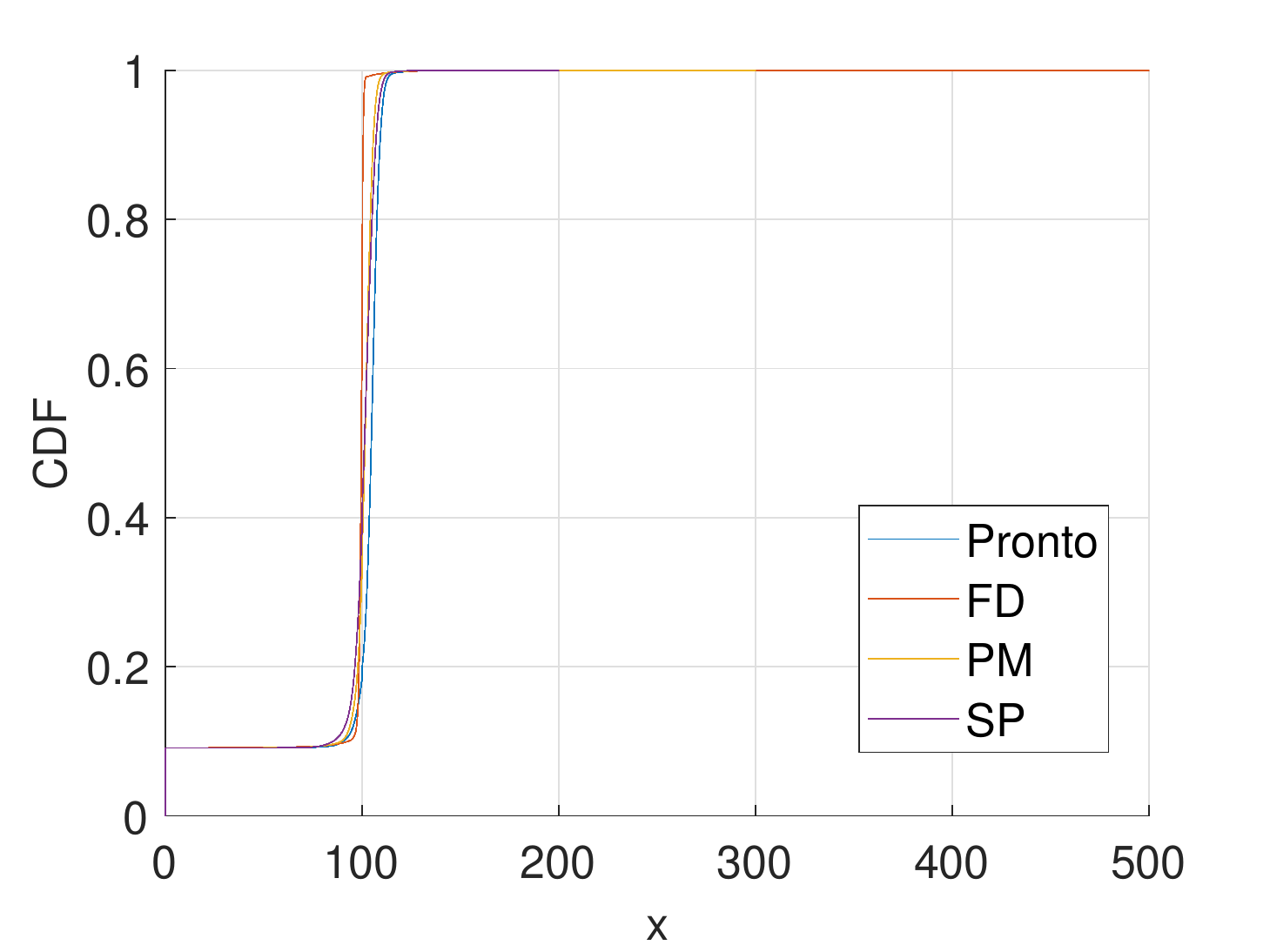}
        \caption{Contained spike percentage CDF}
        \vspace{11pt}
        \label{fig:contained-cdf}
    \end{subfigure}
    \caption{
        Left: The empirical CDF of the percentage of the contained spikes, as we can see all methods perform desirably containing almost all spikes. 
        We note that, values above $100$\% indicate that methods detected more spikes than the ones detected by just using \CPUR. 
        Right: Shows the empirical CDF of the \textit{downtime} of the rejection signal per method, indicating the percentage of the total time a compute job is able to accept an incoming job.
    }
\end{figure}

Figures show that the \textit{left-sided} spikes are considerably more frequent than the \textit{right-sided} ones. This means that we are able to detect incoming \CPUR spikes with high accuracy.
Notably \Pronto and FD have the highest number of \textit{left-sided} spikes, followed by PM and SP. 
Although our goal is to predict \CPUR spikes as accurately as possible we also want to have the highest availability (i.e., nodes \textit{can} accept jobs) for a higher overall utilization.
To quantify this we need to show the CDFs of the overall downtime and the contained spike percentages across the traces; this is shown in~\Cref{fig:downtime-cdf} and~\Cref{fig:contained-cdf} for the downtime and contained spike percentages respectively.

These figures indicate that \Pronto, SPIRIT, and PM have very little downtime compared to the actual spikes detected and thus are able to utilize each compute node more efficiently than FD. 
Notably, we observe that FD performs poorly as its downtime is greater than half of the total time - which means it could be similar to using a random scheduler.
The contained spike percentage CDF shows the amount of total spikes that are contained by all methods considered in this evaluation. 
In this context, values greater than $100\%$ show the proportion of the spikes detected over the actual \CPUR signal spikes.
This could either indicate that these methods detect more congestion points or overestimate and raise the rejection signal without having to do so.
Most methods are always near or over $100\%$ and the skewing present in the graph is mostly attributed to FD.
However, this is to be expected as it is also reflected by the downtime percentage CDF, that indicates FD has almost constantly the rejection signal raised - thus accepting minimal jobs.

\subsection{Performance}

Another important consideration is the performance of each method. 
Our prototype implementation is developed using \texttt{Python} employing standard libraries and established packages (e.g. \texttt{scipy}, \texttt{numpy}).
To this end we measure the performance to update the predictions for each incoming data vector.
Note, that, even if the embedding is updated \textit{per block} the actual predictions job acceptance happens \textit{per data vector}.
We amortize the cost of block methods by averaging the running time over each block to extrapolate the cost to perform the rejection signal computation per incoming data vector.
For memory we opt to use the maximum amount per block or per vector required.
However due to the built-in methods use of ``slack'' space all costs end up being fairly similar.
Further, to avoid confusion we report rounded numbers.
The results for all methods considered are shown in~\Cref{tab:perf_table}. 

\begin{table}[htb!]
    \centering
    \small
     \caption{Extrapolated average execution time (ms) per timestep (\textit{i.e.}, per vector that the rejection signal is computed) and approximate mean memory allocation to process each block/vector, including any overheads (MB).\\}
    \begin{tabular}{l|ll}
          & Execution time & Memory allocation (rounded)  \\
         \hline
         $\Pronto$ & $ \textbf{15}$ ms & $\approx\textbf{148}$ MB \\
         \hline
         PM & $ \textbf{22}$ ms & $\approx\textbf{155}$ MB \\
         FD & $ \textbf{25}$ ms & $\approx\textbf{151}$ MB \\
         SP & $ \textbf{9}$ ms & $\approx\textbf{123}$ MB \\
         \hline
    \end{tabular}
     \label{tab:perf_table}
\end{table}

All of the considered algorithms do not consume a lot of resources and predictions happen in near real-time as they are updated every second (or around $100$ms).
However, we note that even with very small matrices when using built-in functions \texttt{splinalg.svds} and \texttt{linalg.svd} of \texttt{scipy} package the required memory was about $\approx100$ MB at its peak without increasing much even for larger ones. 
That is, we observed that the memory difference for performing $\SVD$ for $40\times 7$ and $300 \times 7$, was $\approx0.45$MB as measured using the \texttt{memory\_profiler} module.
We conjecture that this is due to the use of generous slack space and debug symbols used by \texttt{Python}. 
However during our experiments memory consumption never reached over $\approx150$ MB.
Practically speaking, the most expensive operation throughout the whole pipeline is the computation of $\SVD$.
Though, given the fact that since all methods are able to exploit the truncated $\SVD$ instead of the normal one we are able to keep resource runtime requirements reasonable and able to update the estimates using either per vector or per small blocks.
Finally, we note that the measurements are taken with debug symbols enabled and no particular optimizations; thus potential optimizations could help bring the cost further down.

\section{Related Work}
\noindent
\textbf{Data Center Scheduling using low-level, real-time performance data:}

Today’s data centers (especially virtualized) generate a plethora of real-time telemetry performance data consisting of fine-grained time-series measurements such as CPU utilization, memory consumption, and I/O transactions. 
This data uniquely captures real-time behaviour of the data center and can be used for resource predictions. 
Numerous related works highlight the importance to accurately analyze this data for resource prediction and modelling. We further discuss the most relevant works below.

\textbf{VM resource management and prediction.} In the past, there had been a lot of interest in the area of autonomic VM resource management and prediction using \emph{feedback control}, mostly targeting small-scale cluster deployments.

Early works~\citep{utilizationAndSLO, utilityDriven, adaptiveControlV} use non-linear controllers to regulate the relative CPU utilization across VMs under contention. 
The problem of resource contention in shared virtualized clusters is also studied by \citep{optimalMultivariate}; their approach allocates CPU resources based on a response time ratio between workloads on saturated VMs.
For instance, \citep{automatedControl} use a second-order ARMA model to capture the relationship between resource utilization and application performance. 
\citep{vATM} address the problem of boot storms when multiple VM workloads start to execute at the same time. 
They propose a feedback approach to control the concurrency level, improving end-to-end latency. 
\citep{kalman, kalmanTAAS} propose linear feedback controllers that are based on the Kalman filtering technique and a widely applicable model of CPU utilization. 
These controllers use online measurements to predict future demands and configure their parameters.  
The use of the min-max $\mathcal{H}_\infty$ filters to minimize the maximum error during under-provisioning was explored in~\citep{hinfinity}.  
Although the above approaches use resource utilization data to derive performance models in a real-time manner they however target small-scale virtualized clusters.

The previous feedback control-based approaches ignore regular long-term patterns, which may exist in the time-series data and could potentially assist with performance management issues. 
For example,~\citep{predictiveControl} present a predictive controller that regulates the relative utilization of a single-tier virtualised server based on three time-series prediction algorithms (AR auto-regressive model, the ANOVA
decomposition and the MP multi-pulse model). 
Their results show that, when utilization exhibits regular patterns, their predictive controller outperforms the relative-utilisation feedback controller proposed by~\citep{utilizationAndSLO}.

Additionally, a wavelet-based approach was proposed by~\citep{agile} for online demand prediction of resource utilizations. 
The advantage of such an approach is the decomposition of the original signal into multiple detailed signals that capture different patterns and finally are synthesized into an approximation signal for predictions.
Their approach uses a history of the past several minutes to generate a new model. 
To make short-term predictions in the absence of long-term patterns, PRESS by~\citep{pressWilkes} combines state or signature-based pattern predictors for resource allocation using previous utilization measurements.
CloudScale by~\citep{cloudScale} targets the problem of fast reaction to hotspots in deployments. 
It combines online adaptive padding based on burst detection with additional allocation correction using feedback from SLO violations and relative utilization. 
While this works efficiently for patterns reported in training data, it is unclear if CloudScale can handle scaling of multiple metrics and application with multiple components with correlated utilisation.

The above predictive approaches show that resource utilization predictions are possible when considering long-term history of a measured metric and that this could assist towards resource management problems. 
However, these approaches have either focused on short-term prediction with a small history or long-term patterns as found in a single or small group of monitored resource utilization traces. 
More recent works recognize the need to analyze telemetry data in the large-scale. Most notably, Microsoft’s Recourse Central gathers all VM telemetry data on a centralized cluster and it focuses on offline analysis for predictions to tackle servers’ oversubscriptions~\citep{cortez2017resource}. When compared to these works, \Pronto is the first approach to target real-time predictions of the \CPUR spikes using data from large-scale deployments.

Further, there exists different centralized and distributed approaches in the area of large-scale resource management. 

\noindent \textbf{Centralized schedulers}: %
Centralized schedulers gather performance data from nodes periodically to take globally informed scheduling decisions. 
In doing so, they have a global view of the data center availability and so ultimately they could take well informed, even optimal, allocation decisions under specific constraints and performance goals, e.g., Decima~\citep{decima}, Firmament~\citep{firmament}, Quincy~\citep{quincy}, Resource Central~\citep{cortez2017resource}, Google’s Omega~\citep{schwarzkopf2013omega}, and Tetrisched~\citep{tetrisched}. 
For example, Google’s Borg centralized scheduler provisions for thousands of machines per cell by employing a similar to Omega’s cached state~\citep{schwarzkopf2013omega} and a number of heuristics such as score caching and relaxed randomisation~\citep{borg}. 
More recently, Microsoft’s Resource Central uses performance prediction based on workload characteristics and machine learning to increase resource utilization but requires gathering of resource utilization data in centralized locations~\citep{cortez2017resource}. 
Although powerful, centralized schedulers do not scale easily and often employ heuristics that decrease allocations’ efficiency. Their approach increases the overall end-to-end processing and adds traffic to the data center network. 
In fact, related papers do not adequately discuss their approach to maintaining a consistent central view of the data center, e.g.,~\citep{firmament, cortez2017resource, decima}. 
Furthermore, their decisions are based on old, cached data as by the time solutions are found and new allocations are in place, resource utilizations at the server nodes have changed risking to make scheduling decisions obsolete, e.g., Borg~\citep{borg}. 

\noindent \textbf{Distributed approaches}: To handle scalability, fault-tolerance, and speedup scheduling time, different decentralized schedulers have been proposed. Popular systems such as Kubernetes~\citep{kubernetes}, Mesos~\citep{mesos}, Autopilot~\citep{autopilot}, Yarn~\citep{vavilapalli2013apache}, Sparrow~\citep{Oust}, Medea~\citep{medea}, and Apollo~\citep{apollo} provide scalable and fault-tolerant managers for scheduling but they do not work on a global view of the data center. 
Their focus is on engineering robust, practical and scalable frameworks tailored for specific workloads, albeit though with very good performance. 
\Pronto is the first to use unsupervised learning techniques in the context of scheduling and shows that is able to predict saturation indicated by the \CPUR metric and uses several innovations to do so.
Firstly, it relies on a recently introduced incremental method to update the local estimates that reside within each node~\citep{grammenos2019federated}.
These iterates are then exploited by projecting incoming traces onto them, which results in the generation of the projections reflecting the overall trend of each of the Principal Components contained in that iterate. 
Finally, these insights are used to generate a binary rejection signal which unlocks the ability for each node to perform scheduling decisions independently.
To the best of our knowledge, this is the first scheme that exploits projection tracking in the scheduling domain and is able to accurately predict \CPUR spikes in the unsupervised setting within minimal assumptions over the input data.

\section{Discussion \& Conclusions}
\label{sec:discussion}

In this work we introduced \Pronto, a federated scheduling algorithm that is able to accurately and efficiently predict whether accepting an incoming job by a data center node would negatively impact its responsiveness.
Further, we empirically validated that online, unsupervised techniques can be used to provide timely scheduling decisions that can happen within each node.
This is key, as it eliminates the potential latency of having to either constantly or periodically communicate with a master node and/or update a global shared state as seen in previous related work.
Moreover, as is shown through our evaluation such scheduling decisions can be made in less than a second, within each computation node, and with minimal overhead.
Naturally, there can be edge cases in which such methods can perform poorly but overall we think they strike an attractive balance between cost, performance, and convenience.
This is due to inherent costs associated with maintaining and/or processing large amounts of high-dimensional data which could end up being prohibitively expensive.
Notably, this problem is emphasised as systems that are able to output high-dimensional information signals containing a wealth of information about computation nodes are becoming increasingly common; a recent, yet prominent example is Microsoft's Resource Central~\cite{cortez2017resource} which can produce over 500 features but, techniques that are able to exploit such datasets in an online fashion effectively are not readily available.
Therefore, online techniques, such as \Pronto, that are able to capture, summarise, and exploit high-dimensional signals are highly desirable.
Finally, apart from being an effective scheduler \Pronto could also be a competent cluster insights as a monitoring tool, since it is able to effectively summarise the discovered trends within the captured PC's. 
Each of the PC's is a linear combination of features, which would readily indicate which features contribute the most to its explained variance; thus could provide hints and insights for potential performance bottlenecks.
This exciting avenue of research is left for future work.

\section*{Acknowledgements}

This work was supported by The Alan Turing Institute under grants: TU/C/000003 and EP/N510129/1. We would like to thank Victoria Lopez Morales for her work on offline prediction of \CPUR values and spikes.

\bibliography{refs}
\bibliographystyle{mlsys_style}

\vfill
\pagebreak
\onecolumn
\appendix

\section{Supplementary material}

For completeness, the full derivations and algorithm definitions which are used in the paper but are not direct contributions of this work are shown below.

\subsection{Local Embedding Updates: Merging subspaces}

The algorithmic construction that is used for merging subspaces by~\cite{vrehuuvrek2011subspace,grammenos2019federated} which can be used if we know a-priori that $\mathbf{V}^{T}$ is not needed and we have knowledge that the subspaces to be merged, namely $\mathbf{U}_{1}$ and $\mathbf{U}_{2}$ are already orthonormal. 
Initially we start by using the direct consequence of the $\SVD$ theorem which can be used to merge two principal subspace as follows:

\begin{algorithm}
    \KwData{
        $\B{U}_{1} \in \R^{d \times r_{1}}$, first subspace\\
        $\B{\Sigma}_{1} \in \R^{r_{1} \times r_{1}}$, first subspace singular values\\
        $\B{U}_{2} \in \R^{d \times r_{2}}$, second subspace\\
        $\B{\Sigma}_{2} \in \R^{r_{2} \times r_{2}}$, second subspace singular values\\
        $r \in [r]$, , the desired rank $r$\\
        $\lambda \in (0, 1)$, forgetting factor\\
        $\lambda_{2} \geq 1$, enhancing factor\\
    }
    \KwResult{
        $\B{U}' \in \R^{d \times r}$, merged subspace\\
        $\B{\Sigma}' \in \R^{r \times r}$, merged singular values
    }
    $[\B{U'}, \B{\Sigma'}, \text{\textasciitilde}] \leftarrow \SVD_{r}([\lambda_{1} \B{U}_{1}\B{\Sigma}_{1}, \lambda_{2} \B{U}_{2}\B{\Sigma}_{2}])$\\ 
    \caption{Basic subspace \texttt{Merge} algorithm}
    \label{algorithm:basic_subspace}
\end{algorithm}

As previously mentioned, \Cref{algorithm:basic_subspace} is a direct consequence of the $\SVD$ with the only addition of a forgetting factor $\lambda$ that is intended to provide a weighting factor during merging.
Recall, that in our particular case we do not require $\B{V}^{T}$, which is computed by default when using $\SVD$; this incurs both computational and memory overheads.
We now show how we can do better in this regard. 
This is done by building a basis $\B{U}'$ for $\operatorname{span}((\B{I} - \B{U_1}\B{U_1}^T)\B{U_2})$ via the QR factorisation and then computing the $\SVD$ decomposition of a matrix $\B{X}$ such that  

\begin{equation}
    [\B{U_{1}}\B{\Sigma_{1}},\B{U_{2}}\B{\Sigma_{2}}] = [\B{U_{1}},\B{U}']\B{X}.
\end{equation}

It is shown in~\citep[Chapter 3]{vrehuuvrek2011subspace} in an analytical derivation that this yields an $\B{X}$ of the form

\begin{equation*}
    \B{X} = 
    \begin{bmatrix} 
        \B{U_{1}^{T}}\B{U_{1}}\B{\Sigma_{1}} &
        \B{U_{1}^{T}}\B{U_{2}}\B{\Sigma_{2}}\\
        \B{U'}^{T}\B{U_{1}} & \B{U'^{T}}\B{U_{2}}\B{\Sigma_{2}}
    \end{bmatrix}
    = 
    \begin{bmatrix} 
        \B{\Sigma_{1}} & \B{U_{1}^{T}}\B{U_{2}}\B{\Sigma_{2}}\\
        0 & \B{R_{p}}\B{\Sigma_{2}}
    \end{bmatrix}
\end{equation*} 

The end result, is the algorithm that is shown in below.

\begin{algorithm}[htb!]
\footnotesize{
    \DontPrintSemicolon
    \SetNoFillComment
    \KwData{$\mbox{Merge}_r( \B{U}_1, \B{\Sigma}_1, \B{U}_2, \B{\Sigma}_2)$\;
        
    $(\B{U}_{1}, \B{\Sigma}_1) \in \R^{d \times r_{1}} \times \R^{r_1 \times r_1}$: First subspace\;
    $(\B{U}_{2}, \B{\Sigma}_2) \in \R^{d \times r_{2}} \times \R^{r_2 \times r_2}$: Second subspace\;
    }
    \KwResult{
        $(\B{U}'', \B{\Sigma}'') \in \R^{d \times r} \times \R^{r \times r}$ merged subspace\;
    }
    $\B{Z} \leftarrow \B{U}^{T}_{1}\B{U}_{2}$\;
    $[\B{Q}, \B{R}] \leftarrow \QR(\B{U}_{2} - \B{U}_{1}\B{Z})$\;
    $[\B{U}',\B{\Sigma}'', \thicksim] \leftarrow \SVD_{r}\bigg(
    \begin{bmatrix} 
        \B{\Sigma}_{1} & \B{Z}\B{\Sigma}_{2} \\
        0 & \B{R}\B{\Sigma}_{2}
    \end{bmatrix}
    \bigg)$\;
    $\B{U}'' \leftarrow [\B{U}_{1}, \B{Q}]\B{U}'$\;
    }
    \caption{Merge \citep{vrehuuvrek2011subspace,grammenos2019federated}}
    \label{algorithm:fastest_subspace}
\end{algorithm}

\subsection{Local Embedding Updates: Subspace tracking}
\label{streaming_pca}

Consider a sequence $\{\mathbf{y}_1, \dots, \mathbf{y}_n\} \subset \R^{d}$ of feature vectors.
A block of size $b \in \N$ is formed by taking $b$ contiguous columns of $\{\mathbf{y}_1, \dots, \mathbf{y}_n\}$.
Assume $r \leq b \leq \tau \leq n$. If $\widehat{\Y}_{0}$ is the empty matrix, the $r$ principal components of $\Y_{\tau} := [\mathbf{y}_1, \cdots, \mathbf{y}_{\tau}]$ can be estimated by running the following iteration for $k = \{1, \dots, \lceil \tau/b \rceil\}$,

\begin{equation}
\label{eq-apx:svdr_2}
[\widehat{\B{U}}, \widehat{\B{\Sigma}},\widehat{\B{V}}^T]\leftarrow\text{SVD}_{r}\left(\left[\begin{array}{cccc}
\widehat{\Y}_{(k-1)b} & \mathbf{y}_{(k-1)b+1} & \cdots & \mathbf{y}_{kb}\end{array}\right]\right), \;\;\;\; \widehat{\Y}_{kb}\leftarrow \widehat{\B{U}} \widehat{\B{\Sigma}}\widehat{\B{V}}^T \in \R^{d \times kb}.
\end{equation}

Its output after $K = \lceil \tau /b \rceil$ iterations contains an estimate $\widehat{\B{U}}$ of the leading $r$ principal components of $\Y_{\tau}$ and the projection $\widehat{\Y}_{\tau}=\widehat{\B{U}} \widehat{\B{\Sigma}}\widehat{\B{V}}^T$  of $\Y_{\tau}$ onto this estimate.
The local subspace estimation in \eqref{eq-apx:svdr_2} by~\cite{grammenos2019federated}. 
$\FPCAEC$ adapts \eqref{eq-apx:svdr_2} to the federated setting by implementing an adaptive rank-estimation procedure which allows clients to adjust, independently of each other, their rank estimate based on the distribution of the data seen so far.
That is, by enforcing,
\begin{equation}
\label{eq:sv_energy}
     \energy_r({\Y_{\tau}}) = \frac{\sigma_{r}(\Y_{\tau})}{\sum_{i=1}^{r}\sigma_{i}(\Y_{\tau})}
     \in [ 
     \alpha, \beta],
\end{equation}
and increasing $r$ whenever $\energy_r(\Y_{\tau}) >\beta$ or decreasing it when $\energy_r(\Y_{\tau}) < \alpha$.
In our algorithm, this adjustment happens only once per block, however for merging we only propagate the changes upwards should the absolute difference of the previous and current subspace estimates differ above a certain threshold; of course more complex variations to this strategy are possible.

Letting $[r+1] = \{1, \dots, r+1\}$, $[r-1] = \{1, \dots, r-1\}$, and $\ind\{\cdot\}\in \{0, 1\}$ be the indicator function, the subspace tracking and rank-estimation procedures in~\Cref{alg:fpca_edge} depend~\Cref{algorithm:fastest_subspace} as well as on the following functions: 
\begin{equation*}
\begin{array}{rl}
\RSPCA_r(\mathbf{D}, \mathbf{U}, \mathbf{\Sigma}) =& 
        \SVD_r (\mathbf{D}) \ind\{ \mathbf{U}\mathbf{\Sigma} = 0\} + \mbox{Merge}_{r}(\mathbf{U}, \mathbf{\Sigma}, \mathbf{D}, \mathbf{I}) \ind\{\mathbf{U}\mathbf{\Sigma} \neq 0 \}\\ \\
\rankupdate_r^{\alpha, \beta}({\mathbf{U}}, {\mathbf{\Sigma}}) =& 
        \left([{\mathbf{U}}, \vec{\mathbf{e}}_{r+1}], \B{\Sigma}_{[r+1]} \right) \ind\{\energy_r({\mathbf{\Sigma}}) > \beta\} + 
        ({\mathbf{U}}_{[r-1]}, {\mathbf{\Sigma}}_{[r-1]}) \ind\{\energy_r({\mathbf{\Sigma}}) < \alpha\} \\
        &+ 
        (\mathbf{U}, \mathbf{\Sigma}) \ind\{\energy_r({\mathbf{\Sigma}}) \in [\alpha, \beta]\}
\end{array}
\end{equation*}

\begin{algorithm}[htb!]
\footnotesize{
    \DontPrintSemicolon
    \SetNoFillComment
    \KwData{
    $\FPCAEC_{\alpha, \beta, r}(\B{B}, \wh{\mathbf{U}}_{k-1}, \wh{\mathbf{\Sigma}}_{k-1})$
    
    $\B{B}\in \R^{d \times b}$: {\em Batch $\Y_{\{(k-1)b+1, \dots, kb\}}$}\; 
    $(\wh{\mathbf{U}}_{k-1}, \wh{\mathbf{\Sigma}}_{k-1})$: {\em SVD estimate for $\Y_{\{1, \dots, (k-1)b\}}$}\;
    $r$: {\em Initial rank estimate}\;
    $(\alpha, \beta)$: {\em Bounds on energy, see \eqref{eq:sv_energy}}\; 
    $r$: {\em Initial rank estimate}\;
    }
    \KwResult{ $(\wh{\mathbf{U}},\wh{\mathbf{\Sigma}})$, principal $r$-subspace of $\Y_{\{1, \dots, kb\}}$.}
    \tcc{Update embedding estimates}
    $(\B{U}, \B{\Sigma}) \leftarrow  \RSPCA_r(\B{B}_{s}, \B{U}, \B{\Sigma})$\;
    \tcc{Merge with previous estimate}
    $(\wh{\mathbf{U}}', \wh{\mathbf{\Sigma}}') \leftarrow \mbox{Merge}(\mathbf{U}, \mathbf{\Sigma}, \wh{\mathbf{U}}_{k-1}, \wh{\mathbf{\Sigma}}_{k-1})$\;
    \tcc{Adjust the rank}
    $(\wh{\mathbf{U}}, \wh{\mathbf{\Sigma}}) \leftarrow \rankupdate_r^{\alpha, \beta}(\wh{\mathbf{U}}',\wh{\mathbf{\Sigma}}')$
        }
    \caption{Federated PCA Edge ($\FPCAEC$)}
    \label{alg:fpca_edge} 
\end{algorithm}

\end{document}